\documentstyle[12pt,epsfig,axodraw]{article}
\def\rp{$R_p \hspace{-1em}/\;\:$ }
\def\half{{\textstyle{1 \over 2}}} 
\def\quarter{{\textstyle{1 \over 4}}}

\def\21{$SU(2) \otimes U(1) $}
\def\np#1#2#3{           {Nucl.~Phys.~}{\bf #1} (19#2) #3}
\def\pl#1#2#3{           {Phys.~Lett.~}{\bf #1} (19#2) #3}
\def\pr#1#2#3{           {Phys.~Rev.~}{\bf #1} (19#2) #3}

\def\half{{\textstyle{1 \over 2}}} 
\def\s#1{\tilde{#1}}
\def\lsim{\raise0.3ex\hbox{$\;<$\kern-0.75em\raise-1.1ex\hbox{$\sim\;$}}}
\def\gsim{\raise0.3ex\hbox{$\;>$\kern-0.75em\raise-1.1ex\hbox{$\sim\;$}}}

\newcommand{\wt}{\widetilde}
\newcommand {\chim} [1] {\tilde{\chi}^{-}_{#1} }
\newcommand {\chipm} [1] {\tilde{\chi}^{\pm}_{#1} }
\newcommand {\chiz} [1] {\tilde{\chi}^{0}_{#1} }
\newcommand {\beq} {\begin{equation}}
\newcommand {\eeq} {\end{equation}}
\newcommand {\bea} {\begin{eqnarray}}
\newcommand {\eea} {\end{eqnarray}}
\newcommand {\fig} [1] {Fig.~\ref{#1}}
\newcommand {\figs} [2] {Fig.~\ref{#1} and \ref{#2}}
\newcommand {\eq} [1] {Eq.~(\ref{#1})}
\newcommand {\mchiz} [1] {m_{\tilde{\chi}^{0}_{#1} } }
\newcommand {\missPT} {$p_T \hspace{-4mm} /$ \hspace{0.3mm}}
\newcommand{\mx}{\left[\begin{array}} 
\newcommand{\finmx}{\end{array}\right]} 
\def\nn{\nonumber}
\newcommand {\tab} [1] {Table~\ref{#1} }
\newcommand {\tanb} {\tan \beta}
\begin{document}
\title{
\begin{flushright} \small \rm
  hep-ph/0007157 \\ 
  IFIC/00-45 \\
  UWThPh-2000/23 \\[1cm]
\end{flushright}
Neutralino Phenomenology at LEP2 in Supersymmetry with
Bilinear Breaking of R-parity }
\author{A.~Bartl$^1$, W.~Porod$^2$, 
D.~Restrepo$^2$, J.~Rom\~ao$^3$, and J.~W.~F.~Valle$^2$ \\[0.5cm]
\small $^1$Inst.~f.~Theor.~Physik, Universit\"at Wien,
           A-1090 Vienna, Austria  \\
\small $^2$Inst.~de F\'\i sica Corpuscular (IFIC), CSIC - U. de Val\`encia,
\\ \small Edificio Institutos de Paterna, Apartado de Correos 22085\\
\small E-46071--Val\`encia, Spain \\
\small$^3$ Departamento de F\'\i sica, Instituto Superior T\'ecnico\\
\small     Av. Rovisco Pais 1, $\:\:$ 1049-001 Lisboa, Portugal 
} \maketitle

\begin{abstract}
  We discuss the phenomenology of the lightest neutralino in models
  where an effective bilinear term in the superpotential parametrizes
  the explicit breaking of R-parity.  We consider supergravity
  scenarios where the lightest supersymmetric particle (LSP) is the
  lightest neutralino and which can be explored at LEP2.  We present a
  detailed study of the LSP decay properties and general features of
  the corresponding signals expected at LEP2. We also contrast our
  model with gauge mediated supersymmetry breaking.
\end{abstract}

\section{Introduction}

The search for supersymmetry (SUSY) plays an important role in the
experimental programs of existing high energy colliders like LEP2,
HERA and the Tevatron. It will play an even more important role at
future colliders like LHC or a linear $e^+ e^-$ collider. So far
most of the effort in searching for supersymmetric signatures has been
confined to the framework of R--parity-conserving \cite{phen1}
realizations.
Recent data on solar and atmospheric neutrinos strongly indicate the
need for neutrino conversions~\cite{MSW99,atm99}. Motivated by this
there has been in the last few years a substantial interest in
R-parity violating models~\cite{nulong5}.
The violation of R-parity could arise explicitly as a residual effect
of some larger unified theory \cite{expl}, or spontaneously, through
nonzero vacuum expectation values (VEV's) for scalar neutrinos
\cite{rpold,ProjectiveMassMatrix}. In the first case there is a large
number of unknown parameters characterizing the superpotential of
these models, so that for simplicity these effects are usually studied
assuming in an \emph{ad hoc} way that only a few dominant terms break
R-parity explicitly, usually only one.

We prefer theoretical scenarios which break R--parity only as a result
of the properties of the vacuum~\cite{rprev:1996}.
There are two generic cases of spontaneous R-parity breaking models.
In the first case lepton number is part of the gauge symmetry and
there is a new gauge boson $Z^\prime$ which gets mass via the Higgs
mechanism \cite{ZR}. In this model the lightest SUSY particle (LSP) is
in general a neutralino which decays, therefore breaking R-parity.
The LSP decays mostly to visible states such as
\beq
\label{vis}
\chiz{1} \to f \bar{f} \nu,
\eeq
where $f$ denotes a charged fermion. These decays are mediated by the
$Z$-boson or by the exchange of scalars. 
In the second class of models there appears a physical massless
Nambu-Goldstone boson, called majoron.
The latter arises in \21 models where the breaking of R-parity occurs
spontaneously. In this case {\sl the majoron is the LSP}, which is
stable because it is massless (or nearly so). It leads to an
additional invisible decay mode $\chiz{1} \to \nu + J$, which is
R-parity conserving since the majoron has a large R-odd singlet
sneutrino component \cite{MASIpot3,MASI}.  This decay is absent if
lepton number is gauged, as the majoron is eaten up by a massive
additional Z boson.
 
Although models with spontaneous R-parity breaking
\cite{ZR,MASIpot3,MASI} usually contain additional fields not present
in the MSSM in order to drive the violation of R-parity (expected to
lie in the TeV range), they are characterized by much fewer parameters
than models with explicit breaking of R--parity. 
Most phenomenological features of these models are reproduced by
adding three explicit bilinear R-parity breaking terms to the MSSM
superpotential~\cite{bilrev:1998}.  This renders a systematic way to
study R-parity breaking signals \cite{chitau,top,Vissani} and leads to
effects that can be large enough to be experimentally observable, even
in the case where neutrino masses are as small as indicated by the
simplest interpretation of solar and atmospheric neutrino data
\cite{nulong5}.  Moreover, R-parity violating interactions follow a
specific \emph{pattern } which can be easily characterized. These
features have been exploited in order to describe the R-parity
violating signals expected for chargino production at LEP II
\cite{deCampos:1999mf}.

Here we consider the phenomenology of the lightest neutralino in the
simplest and well motivated class of models with an effective explicit
R-parity breaking characterized by a single bilinear superpotential
term \cite{Diaz:1998xc}.
Apart from the absence of the majoron-emitting $\chiz{1}$ decays
(which are absent in majoron-less models with spontaneous breaking of
R-parity) this bilinear model mimics all features of neutralino decay
properties relevant for our analysis. For simplicity and for
definiteness we consider supergravity scenarios where the lightest
supersymmetric particle (LSP) is the lightest neutralino. We present a
detailed study of the LSP decay properties and general features of the
corresponding signals expected at LEP2.
In the following we denote the minimal SUGRA scenario with conserved
R-parity by mSUGRA.
It is well known that in models with gauge mediated supersymmetry
breaking (GMSB) the lightest neutralino decays \cite{Giudice:1999bp,GMSB1}, 
because
the gravitino is the LSP. We therefore also discuss the possibilities
to distinguish between GMSB and our R-parity breaking model.

\section{The model}
\label{sec:model}

Here we will adopt a supersymmetric Lagrangian specified by the
following superpotential 
\begin{equation}  
W=\varepsilon_{ab}\left[ 
 h_U^{ij}\widehat Q_i^a\widehat U_j\widehat H_2^b 
+h_D^{ij}\widehat Q_i^b\widehat D_j\widehat H_1^a 
+h_E^{ij}\widehat L_i^b\widehat R_j\widehat H_1^a 
-\mu\widehat H_1^a\widehat H_2^b 
\right] + \varepsilon_{ab}\epsilon_i\widehat L_i^a\widehat H_2^b\,,
\label{eq:Wsuppot} 
\end{equation}
where $i,j=1,2,3$ are generation indices, $a,b=1,2$ are $SU(2)$
indices, and $\varepsilon$ is a completely anti-symmetric $2\times2$
matrix, with $\varepsilon_{12}=1$. The symbol ``hat'' over each letter
indicates a superfield, with $\widehat Q_i$, $\widehat L_i$, $\widehat
H_1$, and $\widehat H_2$ being $SU(2)$ doublets with hypercharges
$1/3$, $-1$, $-1$, and $1$ respectively, and $\widehat U$,
$\widehat D$, and $\widehat R$ being $SU(2)$ singlets with
hypercharges $-{\textstyle{4\over 3}}$, ${\textstyle{2\over 3}}$, and
$2$ respectively. The couplings $h_U$, $h_D$, and $h_E$ are $3\times 3$
Yukawa matrices, and $\mu$ and $\epsilon_i$ are parameters with units
of mass.
 
Supersymmetry breaking is parametrized by the standard set of soft
supersymmetry breaking terms 
\begin{eqnarray} 
V_{soft}&=& 
M_Q^{ij2}\widetilde Q^{a*}_i\widetilde Q^a_j+M_U^{ij2} 
\widetilde U^*_i\widetilde U_j+M_D^{ij2}\widetilde D^*_i 
\widetilde D_j+M_L^{ij2}\widetilde L^{a*}_i\widetilde L^a_j+ 
M_R^{ij2}\widetilde R^*_i\widetilde R_j \nonumber\\ 
&&\!\!\!\!+m_{H_1}^2 H^{a*}_1 H^a_1+m_{H_2}^2 H^{a*}_2 H^a_2\nn\\
&&\!\!\!\!- \left[\half M_3\lambda_3\lambda_3+\half M\lambda_2\lambda_2 
+\half M'\lambda_1\lambda_1+h.c.\right] 
\nn\\ 
&&\!\!\!\!+\varepsilon_{ab}\left[ 
A_U^{ij}h_U^{ij}\widetilde Q_i^a\widetilde U_j H_2^b 
+A_D^{ij}h_D^{ij}\widetilde Q_i^b\widetilde D_j H_1^a 
+A_E^{ij}h_E^{ij}\widetilde L_i^b\widetilde R_j H_1^a\right.
\nn\\ 
&&\!\!\!\!\left.-B\mu H_1^a H_2^b+B_i\epsilon_i\widetilde L_i^a H_2^b\right] 
\,,\label{eq:Vsoft}
\end{eqnarray} 

Note that, in the presence of soft supersymmetry breaking terms the
bilinear terms $\epsilon_i$ can not be rotated away, since the
rotation, that eliminates it, re-introduces an R--parity violating
trilinear term, as well as a sneutrino vacuum expectation
value~\cite{Diaz:1998xc}. This happens even in the case where
universal boundary conditions are adopted for the soft breaking terms
at the unification scale, since universality will be effectively
broken at the weak scale due to calculable renormalization effects.
For definiteness and simplicity we will adopt this assumption
throughout this paper.

Although for the discussion of flavour--changing processes, such as
neutrino oscillations involving all three generations, it is important
to consider the full three-generation structure of the model, for the
following discussion of neutralino decay properties it will suffice to
assume R-parity Violation (RPV) only in the third generation, as a
first approximation.  In this case we will omit the labels $i,j$ in
the superptotential and the soft breaking terms
\cite{Diaz:1998xc,otros}
\begin{eqnarray} 
W&=&h_t\widehat Q_3\widehat U_3\widehat H_2
 +h_b\widehat Q_3\widehat D_3\widehat H_1
 +h_{\tau}\widehat L_3\widehat R_3\widehat H_1
 -\mu\widehat H_1\widehat H_2
 +\epsilon_3\widehat L_3\widehat H_2
\label{eq:Wbil} \\
V_{soft}&=&  \varepsilon_{ab}\left[ 
A_t h_t\widetilde Q_3^a\widetilde U_3 H_2^b 
+A_b h_b \widetilde Q_3^b\widetilde D_3 H_1^a 
+A_\tau h_\tau\widetilde L_3^b\widetilde R_3 H_1^a\right.
\nn\\ 
&&\!\!\!\!\left.-B\mu H_1^a H_2^b+B_3\epsilon_3\widetilde L_3^a H_2^b\right] 
\, + \mathrm{mass \,\, terms} .
\label{eq:Vsoft2}
\end{eqnarray}
This amounts to neglecting the \rp effects in the two first families.

The bilinear terms in Eqs.~(\ref{eq:Wbil}) and (\ref{eq:Vsoft2}) lead
to a mixing between the charginos and the $\tau$--lepton which is
described by the mass matrix
\begin{equation}
{\bf M_C}=\left[\matrix{
M_2 & {\textstyle{1\over{\sqrt{2}}}}gv_2 & 0 \cr
{\textstyle{1\over{\sqrt{2}}}}gv_1 & \mu & 
-{\textstyle{1\over{\sqrt{2}}}}h_{\tau}v_3 \cr
{\textstyle{1\over{\sqrt{2}}}}gv_3 & -\epsilon_3 &
{\textstyle{1\over{\sqrt{2}}}}h_{\tau}v_1}\right]
\label{eq:ChaM6x6} \, ,
\end{equation}
where $v_1$, $v_2$, and $v_3$ are the vevs of $H^0_1$, $H^0_2$, and
$\tilde \nu_\tau$, respectively.  As in the MSSM, the chargino mass
matrix is diagonalized by two rotation matrices $\bf U$ and $\bf V$
\begin{equation}
{\bf U}^*{\bf M_C}{\bf V}^{-1}=\left[\matrix{
m_{\s\chi^{\pm}_1} & 0 & 0 \cr
0 & m_{\s\chi^{\pm}_2} & 0 \cr
0 & 0 & m_{\tau}}\right]\,.
\label{eq:ChaM3x3}
\end{equation}
The lightest eigenstate of this mass matrix must be the tau lepton
($\tau^{\pm}$) and so the mass is constrained to be 1.7771 GeV.  As
explained in~\cite{Akeroyd:1998iq}, the tau Yukawa coupling becomes a
function of the SUSY parameters appearing in the mass matrix.

The neutralino mass matrix is given by:
\beq
{\bf M_N}=\left[\matrix{
M_1 & 0 & -g_1 v_1  & g_1 v_2 & -g_1 v_3 \cr
 0 & M_2 & g_2 v_1 & -g_2 v_2 & g_2 v_3\cr
 -g_1 v_1 & g_2 v_1 & 0 & - \mu & 0\cr
 g_1 v_2 & -g_2 v_2 & - \mu &  0 & \epsilon_3 \cr
 -g_1 v_3 & g_2 v_3 & 0 & \epsilon_3 & 0} \right]\,,
\label{nino}
\eeq 
where $g_1 = g'/2$ and $g_2 = g/2$ denote gauge couplings. This matrix
is diagonalised by a $5 \times 5$ unitary matrix N,
\beq
{\chi}_i^0 = N_{ij} {\psi}_j^0 ,
\eeq
where
$\psi_j^0 = (-i \tilde{B}, -i \tilde{W}_3, \tilde{H}_d,
 \tilde{H}_u, {\nu}_\tau$). 

The up squark mass matrix is given by
\beq
\label{eq:UpSquarkMass}
M_{\tilde u}^2=\left[ \begin{array}{cc}
{M_Q^2}+\frac12 v_2^2 {h_u}^2+\Delta_{UL}&
\frac{h_u}{\sqrt2}
  \left( v_2{A_u}- \mu v_1 +\epsilon_3 v_3 \right) \\
\frac{h_u}{\sqrt2}
  \left( v_2{A_u}- \mu v_1 +\epsilon_3 v_3 \right) &
{M_U^2}+\frac12 v_2^2{h_u}^2+\Delta_{UR}
\end{array} \right]
\eeq
and the down squark mass matrix by
\beq
\label{eq:DownSquarkMass}
M_{\tilde d}^2=\left[ \begin{array}{cc}
{M_Q^2}+\frac12 v_1^2 {h_d}^2+\Delta_{DL}&
\frac{h_d}{\sqrt2}
  \left( v_1{A_d}- \mu v_2  \right) \\
\frac{h_d}{\sqrt2}
  \left( v_1{A_d}- \mu v_2  \right) &
{M_D^2}+\frac12 v_1^2{h_d}^2+\Delta_{DR}
\end{array} \right]
\eeq
with 
$\Delta_{UL}=\frac18\big(g^2-\frac13{g'}^2\big)\big(v_1^2-v_2^2 +v_3^2\big)$,
$\Delta_{DL}=\frac18\big(-g^2-\frac13{g'}^2\big)\big(v_1^2-v_2^2 +v_3^2\big)$,
$\Delta_{UR}=\frac16 {g'}^2(v_1^2-v_2^2+v_3^2)$,
and $\Delta_{DR}=-\frac{1}{12} {g'}^2(v_1^2-v_2^2+v_3^2)$.
The sum of the $v^2_i$ is given by $m^2_W = g^2 (v_1^2+v_2^2+v_3^2)/2$.
The mass eigenstates are given by
$
\wt q_1 = \wt q_L \cos \theta_{\wt q} + 
 \wt q_R \sin \theta_{\wt q}$ and 
$\wt q_2 = \wt q_R \cos \theta_{\wt q} - 
 \wt q_L \sin \theta_{\wt q}.
$
The sfermion mixing angle is given by
\beq
\cos \theta_{\wt q} = \frac{- M^2_{\wt q_{LR}}}{\sqrt{(M^2_{\wt q_{LL}} -
m^2_{\wt q_1})^2 + (M^2_{\wt q_{LR}})^2}}, \qquad 
\sin \theta_{\wt q} = \frac{M^2_{\wt q_{LL}} - m^2_{\wt q_1}}
{\sqrt{(M^2_{\wt q_{LL}} - m^2_{\wt q_1})^2 + (M^2_{\wt q_{LR}})^2}} \, .
\eeq
Here $M^2_{\wt q_{ij}}$ are the corresponding entries of the mass matrices
in Eqs.~(\ref{eq:UpSquarkMass}) and (\ref{eq:DownSquarkMass}).

In addition the charged Higgs bosons mix with charged sleptons and the
real (imaginary) parts of the sneutrino mix the scalar (pseudoscalar)
Higgs bosons. The formulas can be found in \cite{Akeroyd:1998iq,stop3}
and are reproduced, for completeness, in the appendix. The corresponding
mass eigenstates are denoted by $S^+_i$ for the charged scalars, 
$S^0_j$ for the neutral scalars,  and $P^0_k$ for the pseudoscalars. 

\section{Numerical results}
\label{sec:num}

In this section we present numerical predictions for the lightest and
second lightest neutralino production cross sections in $e^+ e^-$
collisions, namely, $e^+ e^-\to \chiz{1} \chiz{1},\chiz{1} \chiz{2}$.
Moreover we will characterize in detail all branching ratios for the
lightest neutralino decays, which violate R-parity.

The relevant parameters include the \rp parameters and
the standard mSUGRA parameters $M_{1/2}$, $m_0$, $\tan \beta$, where
$M_{1/2}$ is the common gaugino mass, $m_0$ the common scalar mass, and
$\tan \beta=v_2/v_1$ is the ratio of the vacuum expectation values of the Higgs
fields.  The absolute value of $\mu$ is fixed by radiative breaking of
electroweak symmetry.  We take $\mu$ positive to be in agreement with
the $b \to s \gamma$ decay \cite{Diaz:1999wq}.  As representative
values of $\tan \beta$ we take $\tan \beta = 3$ and 50.  It is a
feature of models with purely spontaneous breaking of R--parity that
neutrinos acquire a mass only due to the violation of R-parity
\cite{rpold,ProjectiveMassMatrix,mnutreeJ}.
This feature also applies to models characterized by purely bilinear
breaking of R--parity, like our reference model charaterized by
Eqs.~(\ref{eq:Wbil}) and (\ref{eq:Vsoft2}).  As a result the \rp
violating parameters are directly related with $m_{\nu_3}$, the mass
of the neutrino $\nu_3$, which is generated due to the mixing implicit
in \eq{nino}.

\subsection{Neutralino Production}

\begin{figure}
\setlength{\unitlength}{1mm}
\begin{picture}(150,90)
\put(-3,0){\mbox{\epsfig{figure=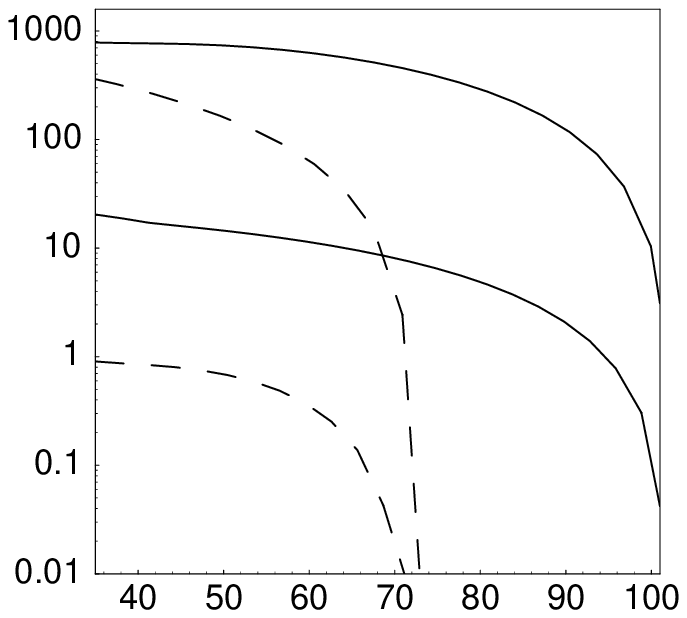,height=8.7cm,width=7.cm}}}
\put(-1,89){{\small \bf a)}}
\put(5,87){\makebox(0,0)[bl]{{\small $\sigma$~[fb]}}}
\put(45,16){\makebox(0,0)[bl]{{\small $\tan\beta =3$}}}
\put(10,44){\makebox(0,0)[bl]{{\small $e^+ e^- \to \chiz{1} \chiz{2}$}}}
\put(34,62){\makebox(0,0)[bl]{{\small $e^+ e^- \to \chiz{1} \chiz{1}$}}}
\put(69,-3){\makebox(0,0)[br]{{ $m_{\chiz{1}}$~[GeV]}}}
\put(73,0){\mbox{\epsfig{figure=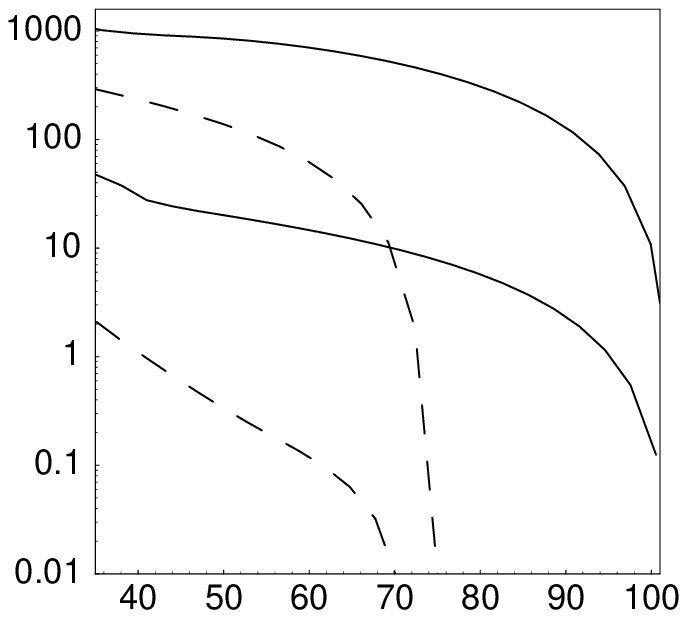,height=8.7cm,width=7.cm}}}
\put(75,89){{\small \bf b)}}
\put(81,87){\makebox(0,0)[bl]{{\small $\sigma$~[fb]}}}
\put(120,18){\makebox(0,0)[bl]{{\small $\tan\beta =50$}}}
\put(87,44){\makebox(0,0)[bl]{{\small $e^+ e^- \to \chiz{1} \chiz{2}$}}}
\put(108,62){\makebox(0,0)[bl]{{\small $e^+ e^- \to \chiz{1} \chiz{1}$}}}
\put(143,-3){\makebox(0,0)[br]{{ $m_{\chiz{1}}$~[GeV]}}}
\end{picture}
\caption[]{Maximum and minimum
  attainable values for the $e^+ e^- \to \chiz{1} \chiz{1}$ (full
  lines) and $e^+ e^- \to \chiz{1} \chiz{2}$ (dashed lines) production
  cross sections in fb as a function of $m_{\chiz{1}}$ for $\sqrt{s} =
  205$~GeV, 50~GeV $<m_0<500$~GeV, 90~GeV $<M_{1/2}<270$~GeV, a)
  $\tan\beta=3$, and b) $\tan\beta=50$.  ISR corrections are
  included.}
\label{fig:ProdMchi}
\end{figure}

While the violation of R--parity would allow for the single production
of supersymmetric particles~\cite{chitau}, for the assumed values of
the \rp violation parameters indicated by the simplest interpretation
of solar and atmospheric neutrino data ~\cite{MSW99,atm99}, these
cross sections are typically too small to be observable.  As a result
neutralino production at LEP2 in our model typically occurs in pairs
with essentially the same cross sections as in the mSUGRA case.
In \fig{fig:ProdMchi}a and b we show the maximum and minimum
attainable values for the $e^+ e^- \to \chiz{1} \chiz{1}$ and $e^+ e^-
\to \chiz{1} \chiz{2}$ production cross sections as a function of
$\mchiz{1}$ at $\sqrt{s}=205$~GeV. We compare the cases $\tanb=3$ and
$\tanb=50$, varying $M_{1/2}$ between 90~GeV and 260~GeV and $m_0$
between 50~GeV and 500~GeV. One can see that, indeed, these results
are identical to those obtained in the mSUGRA.  The $\chiz{1}
\chiz{1}$ production cross section can reach approximately 1~pb.
In our calculation we have used the formula as given in
\cite{Bartl86b} and, in addition, we have included initial state
radiation (ISR) using the formula given in \cite{ISR}. Note that
$\tilde e_{L}$ and $\tilde e_{R}$ are exchanged in the $t$- and
$u$-channel implying that a large fraction of the neutralinos will be
produced in the forward and backward directions.

\begin{figure}
\setlength{\unitlength}{1mm}
\begin{picture}(150,160)
\put(-3,100){\mbox{\epsfig{figure=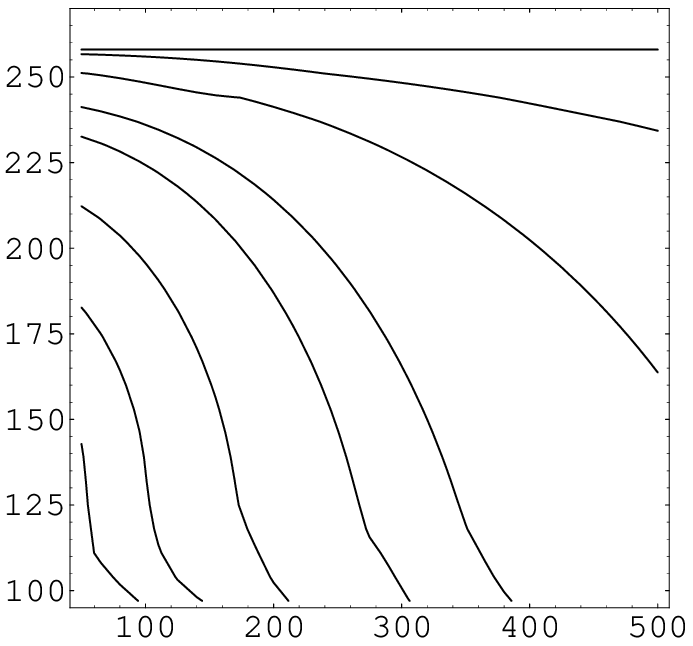,height=8.7cm,width=7.cm}}}
\put(-1,189){{\small \bf a)}}
\put(5,187){\makebox(0,0)[bl]{{\small $M_{1/2}$~[GeV]}}}
\put(29,180){\makebox(0,0)[bl]{{\small $0$}}}
\put(29,172.5){\makebox(0,0)[bl]{{\small $1$}}}
\put(29,165){\makebox(0,0)[bl]{{\small $10$}}}
\put(29,147){\makebox(0,0)[bl]{{\small $50$}}}
\put(25,129){\makebox(0,0)[bl]{{\small $100$}}}
\put(23,120){\makebox(0,0)[bl]{{\small $250$}}}
\put(15,116){\makebox(0,0)[bl]{{\small $500$}}}
\put(7,116){\makebox(0,0)[bl]{{\small $750$}}}
\put(47,130){\makebox(0,0)[bl]{{\small $\tan\beta =3$}}}
\put(69,97){\makebox(0,0)[br]{{ $m_0$~[GeV]}}}
\put(73,100){\mbox{\epsfig{figure=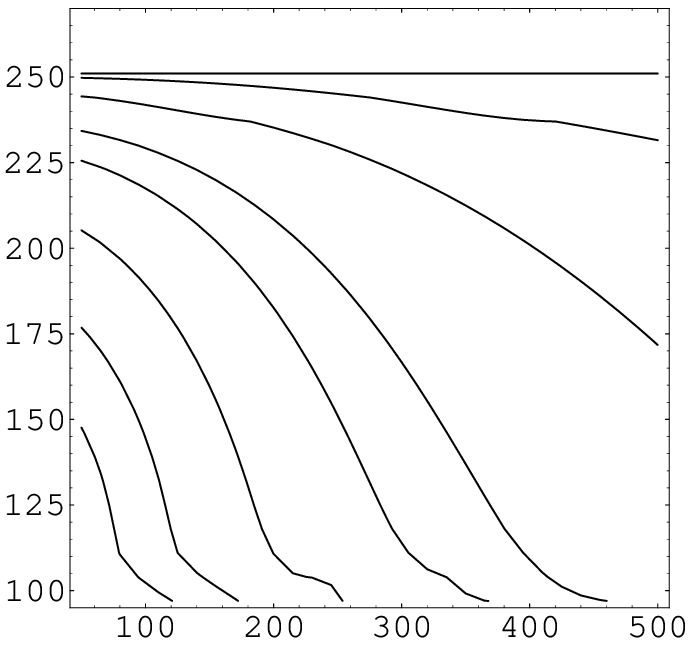
                           ,height=8.7cm,width=7.cm}}}
\put(75,189){{\small \bf b)}}
\put(81,187){\makebox(0,0)[bl]{{\small $M_{1/2}$~[GeV]}}}
\put(122,130){\makebox(0,0)[bl]{{\small $\tan\beta =50$}}}
\put(110,177){\makebox(0,0)[bl]{{\small $0$}}}
\put(110,169){\makebox(0,0)[bl]{{\small $1$}}}
\put(110,160){\makebox(0,0)[bl]{{\small $10$}}}
\put(110.5,146){\makebox(0,0)[bl]{{\small $50$}}}
\put(112,124){\makebox(0,0)[bl]{{\small $100$}}}
\put(100,121){\makebox(0,0)[bl]{{\small $250$}}}
\put(91.5,117){\makebox(0,0)[bl]{{\small $500$}}}
\put(86,113){\makebox(0,0)[bl]{{\small $750$}}}
\put(143,97){\makebox(0,0)[br]{{ $m_0$~[GeV]}}}
\put(-3,0){\mbox{\epsfig{figure=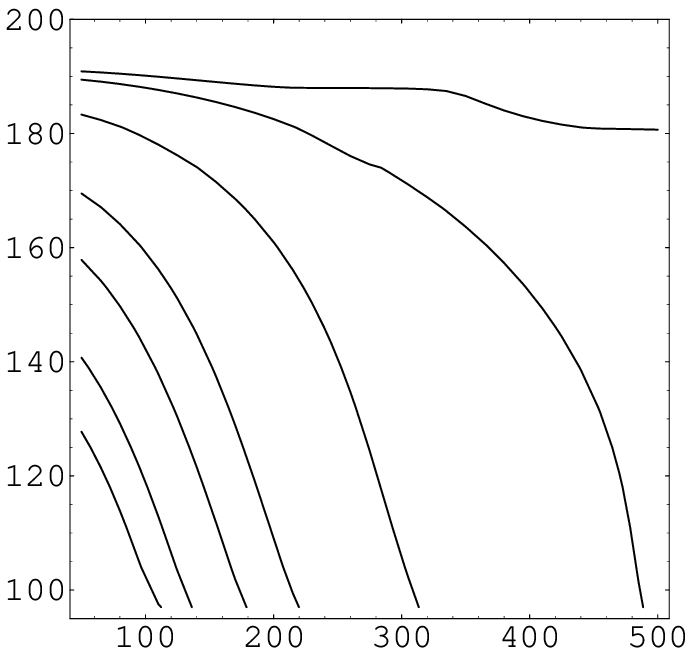,height=8.7cm,width=7.cm}}}
\put(-1,89){{\small \bf c)}}
\put(5,87){\makebox(0,0)[bl]{{\small $M_{1/2}$~[GeV]}}}
\put(29,70){\makebox(0,0)[bl]{{\small $10^{-3}$}}}
\put(27,65){\makebox(0,0)[bl]{{\small $1$}}}
\put(26,55){\makebox(0,0)[bl]{{\small $10$}}}
\put(11,55){\makebox(0,0)[bl]{{\small $50$}}}
\put(12,39){\makebox(0,0)[bl]{{\small $100$}}}
\put(11,25){\makebox(0,0)[bl]{{\small $200$}}}
\put(5,9){\makebox(0,0)[bl]{{\small $300$}}}
\put(25,77){\makebox(0,0)[bl]{{\small $\tan\beta =3$}}}
\put(69,-3){\makebox(0,0)[br]{{ $m_0$~[GeV]}}}
\put(73,0){\mbox{\epsfig{figure=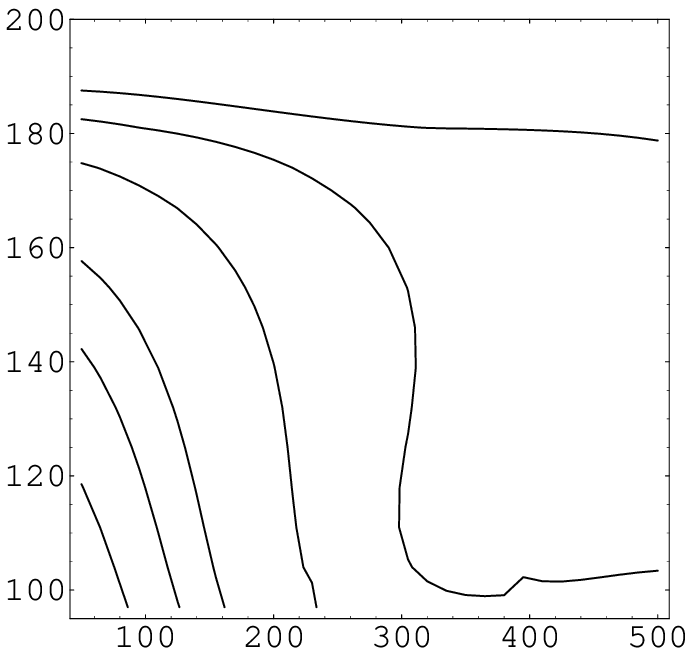,height=8.7cm,width=7.cm}}}
\put(75,89){{\small \bf d)}}
\put(81,87){\makebox(0,0)[bl]{{\small $M_{1/2}$~[GeV]}}}
\put(105,72){\makebox(0,0)[bl]{{\small $\tan\beta =50$}}}
\put(115,66){\makebox(0,0)[bl]{{\small $10^{-3}$}}}
\put(105,64){\makebox(0,0)[bl]{{\small $1$}}}
\put(98,51.5){\makebox(0,0)[bl]{{\small $10$}}}
\put(89,42){\makebox(0,0)[bl]{{\small $50$}}}
\put(81.5,41.5){\makebox(0,0)[bl]{{\small $100$}}}
\put(85,13){\makebox(0,0)[bl]{{\small $200$}}}
\put(143,-3){\makebox(0,0)[br]{{ $m_0$~[GeV]}}}
\end{picture}
\caption[]{
  Contour lines of the production cross sections in fb, in the
  $m_0$--$M_{1/2}$ plane for $\sqrt{s} = 205$~GeV, a) $e^+ e^- \to
  \chiz{1} \chiz{1}$, $\tan\beta=3$, b) $e^+ e^- \to \chiz{1}
  \chiz{1}$, $\tan\beta=50$, c) $e^+ e^- \to \chiz{1} \chiz{2}$,
  $\tanb=3$, and d) $e^+ e^- \to \chiz{1} \chiz{2}$, $\tan\beta=50$.
  ISR corrections are included.}
\label{fig:Prod11Sugra}
\end{figure}

In order to show more explicitly the dependence of the cross sections
on the parameters $m_0$ and $M_{1/2}$ we plot in
\fig{fig:Prod11Sugra}a and b the contour lines of $\sigma(e^+ e^- \to
\chiz{1} \chiz{1})$ in the $m_0$-$M_{1/2}$ plane at $\sqrt{s} =
205$~GeV for $\tanb =3$ and $\tanb = 50$. The contour lines for
$\sigma(e^+ e^- \to \chiz{1} \chiz{2})$ are given in
\fig{fig:Prod11Sugra}c and d.

\subsection{Neutralino Decay Length}

If unprotected by the ad hoc assumption of R--parity conservation the
LSP will decay as a result of gauge boson, squark, slepton and Higgs
boson exchanges.  The relevant contributions to these decays are given
in Table 1.  The Feynman diagrams for the decays not involving taus,
i.e. $\chiz{1} \to \nu_3 f \bar{f}$ ($f=e$, $\nu_e$, $\mu$, $\nu_\mu$,
$u$, $d$, $c$, $s$, $b$) are shown explicitly in \fig{fig:graphs}.
\begin{figure}[h] 
\begin{center}
\begin{picture}(340,160)(0,0) 
\def\punto{2} 
\newcommand{\treebodyuno}[9]{
\def\xx{#1}
\def\largo{#2}           
\def\alto{#3}           
\def\altolabel{#4}      
\def\altotext{#5}       
\def\largophoton{#6}    
\def\altophoton{#7}     
\def\altophotontext{#8} 
\def\largofirst{#9}}      
\newcommand{\treebodydos}[5]{
\def\altofirst{#1}      
\def\largosecond{#2}    
\def\altosecond{#3}     
\def\largothird{#4}     
\def\altothird{#5}  
}      
\newcommand{\treebody}[6]{
\Text(\xx,\altolabel)[]{#1}
\Text(\xx,\altotext)[]{\hspace{45pt}#2}
\ArrowLine(\xx,\alto)(\largo,\alto)
\Vertex(\largo,\alto){\punto}
\Text(\largofirst,\altofirst)[]{\hspace{36pt}#3}
\ArrowLine(\largo,\alto)(\largofirst,\altofirst)
\Text(\largophoton,\altophotontext)[]{\hspace{-56pt}#4}
\Vertex(\largophoton,\altophoton){\punto}
\Text(\largosecond,\altosecond)[]{\hspace{33pt}#5}
\ArrowLine(\largophoton,\altophoton)(\largosecond,\altosecond)
\Text(\largothird,\altothird)[]{\hspace{30pt}#6}
\ArrowLine(\largophoton,\altophoton)(\largothird,\altothird)
}
\treebodyuno{0}{50}{110}{145}{120}{110}{100}{95}{80}
\treebodydos{130}{140}{120}{140}{80}
\treebody{a)}{${\tilde\chi_1^0}(p_1)$}{$\nu_3(p_2)$}{$Z^0$}{$f(p_3)$}{$\bar{f}(p_4)$}
\Photon(\largo,\alto)(\largophoton,\altophoton){2}{8}
\treebodyuno{190}{240}{110}{145}{120}{300}{100}{95}{270}
\treebodydos{130}{330}{120}{330}{80}
\treebody{b)}{${\tilde\chi_1^0}(p_1)$}{$\nu_3(p_2)$}{$S^0_i\,;P^0_i$}{$f(p_3)$}{$\bar{f}(p_4)$}
\DashArrowLine(\largo,\alto)(\largophoton,\altophoton){6}
\treebodyuno{0}{50}{30}{65}{40}{110}{20}{15}{80}
\treebodydos{50}{140}{40}{140}{0}
\treebody{c)}{${\tilde\chi_1^0}(p_1)$}{$f(p_3)$}{$\bar{\!\!\tilde f}_k$}{$\nu_3(p_2)$}{$\bar{f}(p_4)$}
\DashArrowLine(\largo,\alto)(\largophoton,\altophoton){6}
\treebodyuno{190}{240}{30}{65}{40}{300}{20}{15}{270}
\treebodydos{50}{330}{40}{330}{0}
\treebody{d)}{${\tilde\chi_1^0}(p_1)$}{$\bar{f}(p_4)$}{${\tilde f}_k$}{$\nu_3(p_2)$}{$f(p_3)$}
\DashArrowLine(\largo,\alto)(\largophoton,\altophoton){6}
\end{picture}
\end{center}
\caption[]{ Feynman graphs for the decay 
  $\chiz{1} \to \nu_3 \, f \, \bar{f}$ where $f \neq \tau$.}
\label{fig:graphs}
\end{figure}

For the observability of the R-parity violating effects it is crucial
that with this choice of parameters the LSP will decay most of the
time inside the detector.
The neutralino decay path expected at LEP2 depends crucially on the
values of \rp violating parameters or, equivalently, on the value of
the heaviest neutrino mass, $m_{\nu_3}$.
We fix the value of $m_{\nu_3}$ as indicated by the analyses of the
atmospheric neutrino data \cite{atm99}. It is important to note that,
as explained in \cite{nulong5}, due to the projective nature of the
neutrino mass matrix \cite{ProjectiveMassMatrix}, only one of the
three neutrinos picks up a mass in tree approximation. This means
that, neglecting radiative corrections which give small masses to the
first two neutrinos in order to account for the solar neutrino data,
the neutralino decay length scale is set mainly by the tree--level
value of $m_{\nu_3}$. In ref.~\cite{nulong5} we have explicitly shown
that this is a good approximation for most points in parameter space.

\begin{figure}
\setlength{\unitlength}{1mm}
\begin{picture}(150,90)
\put(-1,-4){\mbox{\epsfig{figure=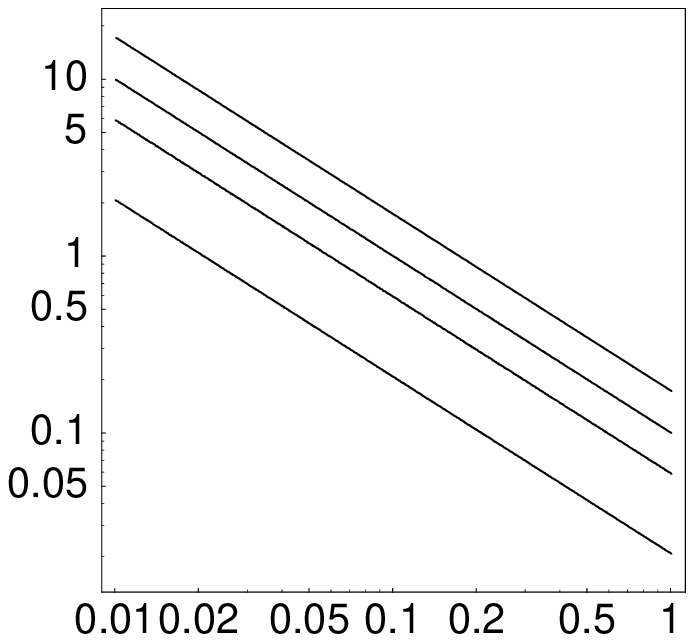,height=9.0cm,width=7.6cm}}}
\put(-1,87){{\small \bf a)}}
\put(5,85){\makebox(0,0)[bl]{{\small $L(\chiz{1})$~[cm]}}}
\put(55,48){\makebox(0,0)[bl]{{\small 60}}}
\put(55,41){\makebox(0,0)[bl]{{\small 70}}}
\put(55,36){\makebox(0,0)[bl]{{\small 80}}}
\put(55,26){\makebox(0,0)[bl]{{\small 90}}}
\put(77,-3){\makebox(0,0)[br]{{ $m_{\nu_3}$~[eV]}}}
\put(82,-6){\mbox{\epsfig{figure=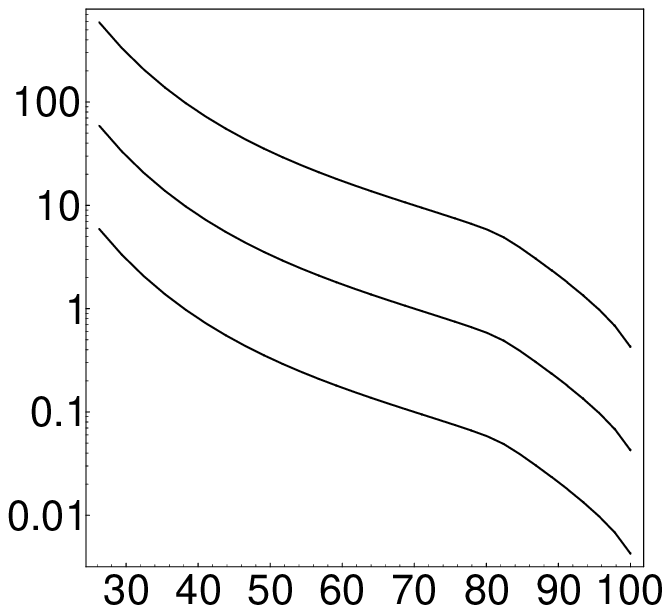,height=9.4cm,width=7.6cm}}}
\put(81,87){{\small \bf b)}}
\put(89,85){\makebox(0,0)[bl]{{\small $L(\chiz{1})$~[cm]}}}
\put(125,60){\makebox(0,0)[bl]{{\small 0.01}}}
\put(125,46){\makebox(0,0)[bl]{{\small 0.1}}}
\put(125,32){\makebox(0,0)[bl]{{\small 1}}}
\put(158,-3){\makebox(0,0)[br]{{ $m_{\chiz{1}}$~[GeV]}}}
\end{picture}
\caption[]{Decay length of the lightest neutralino in cm 
for $\sqrt{s} = 205$~GeV, a) as a function
of $m_{\nu_3}$ for $m_{\chiz{1}} = 60, 70, 80$, and 90 GeV, b) as a function
of $m_{\chiz{1}}$ for $m_{\nu_3} = 0.01, 0.1$, and 1 eV.}
\label{fig:LengthMass}
\end{figure}

\begin{figure}
\setlength{\unitlength}{1mm}
\begin{picture}(150,95)
\put(-3,0){\mbox{\epsfig{figure=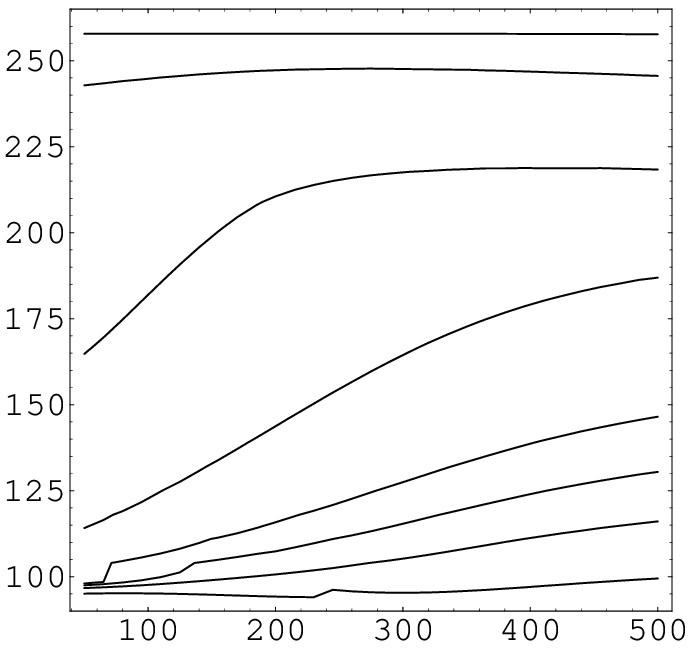,height=8.7cm,width=7.cm}}}
\put(-1,89){{\small \bf a)}}
\put(5,87){\makebox(0,0)[bl]{{\small $M_{1/2}$~[GeV]}}}
\put(38,78){\makebox(0,0)[bl]{{\small $0.001$}}}
\put(26,73){\makebox(0,0)[bl]{{\small $0.1$}}}
\put(26,63){\makebox(0,0)[bl]{{\small $1$}}}
\put(26,36){\makebox(0,0)[bl]{{\small $10$}}}
\put(57,33){\makebox(0,0)[bl]{{\small $50$}}}
\put(57,27){\makebox(0,0)[bl]{{\small $100$}}}
\put(57,20){\makebox(0,0)[bl]{{\small $200$}}}
\put(57,13){\makebox(0,0)[bl]{{\small $500$}}}
\put(30,53){\makebox(0,0)[bl]{{\small $\tan\beta =3$}}}
\put(69,-3){\makebox(0,0)[br]{{ $m_0$~[GeV]}}}
\put(73,0){\mbox{\epsfig{figure=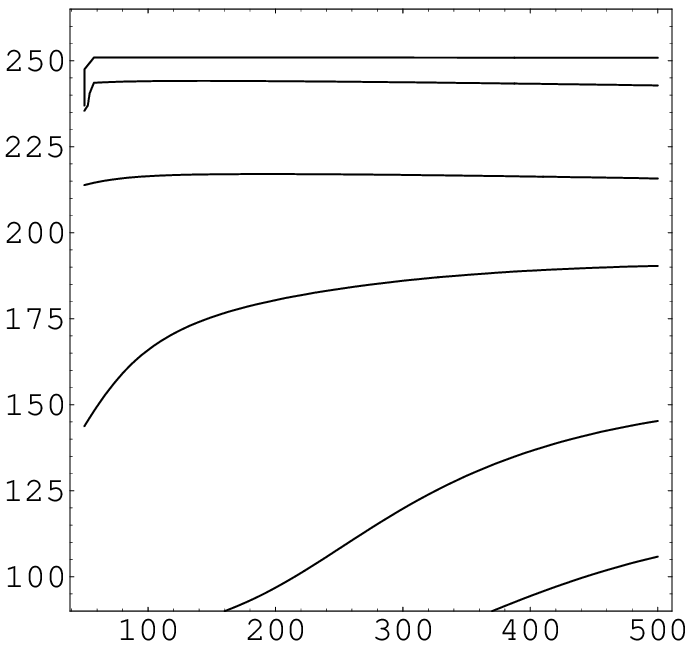,height=8.7cm,width=7.cm}}}
\put(75,89){{\small \bf b)}}
\put(81,87){\makebox(0,0)[bl]{{\small $M_{1/2}$~[GeV]}}}
\put(91,53){\makebox(0,0)[bl]{{\small $\tan\beta =50$}}}
\put(98,79){\makebox(0,0)[bl]{{\small $0.001$}}}
\put(96,72){\makebox(0,0)[bl]{{\small $0.1$}}}
\put(96,64){\makebox(0,0)[bl]{{\small $1$}}}
\put(130,53){\makebox(0,0)[bl]{{\small $10$}}}
\put(130,31){\makebox(0,0)[bl]{{\small $50$}}}
\put(130,14.5){\makebox(0,0)[bl]{{\small $100$}}}
\put(143,-3){\makebox(0,0)[br]{{ $m_0$~[GeV]}}}
\end{picture}
\caption[]{Decay length of the lightest neutralino in cm 
in the $m_0$--$M_{1/2}$ plane for $\sqrt{s} = 205$~GeV, 
a) $\tan\beta=3$, and b)  $\tan\beta=50$. The R-parity violating parameters
are fixed such that $m_{\nu_3} = 0.06$~GeV.}
\label{fig:LengthSugra}
\end{figure}

In \fig{fig:LengthMass} we plot the $\chiz{1}$ decay length in cm
expected at LEP2 for $\sqrt{s} = 205$~GeV. Here and later on
we consider the neutralinos
stemming from the process $e^+ e^- \to \chiz{1} \chiz{1}$ when discussing
the decay length.
In \fig{fig:LengthMass}a we plot the $\chiz{1}$ decay length in cm
as a function of neutrino
mass $m_{\nu_3}$, for different $\mchiz{1}$ between 60 and 90 GeV,
with $m_0 = 100$ ~GeV, and $\tan \beta = 3$. As can be seen the
expected neutralino decay length is typically such that the decays
occur inside the detector, leading to a drastic modification of the
mSUGRA signals.  An equivalent way of presenting the neutralino decay
path at LEP2 is displayed in \fig{fig:LengthMass}b, which gives the
decay length of $\chiz{1}$ as a function of $\mchiz{1}$ for $m_{\nu_3}
= 0.01$, 0.1, and 1 eV.
Finally, we show the dependence of the neutralino decay path on the
supergravity parameters fixing the magnitude of \rp violating
parameters or, equivalently, the magnitude of the heaviest neutrino
mass, $m_{\nu_3}$.
In \fig{fig:LengthSugra}a and b we plot the contour lines of the decay
length of $\chiz{1}$ in the $m_0$-$M_{1/2}$ plane for $m_{\nu_3} =
0.06$~eV, $\tanb =3$ and 50. Note that the decay length is short
enough that it may happen inside typical high energy collider
detectors even for the small neutrino mass values $\sim 0.06$~eV
indicated by the atmospheric neutrino data~\cite{atm99}.
For large values of $\tanb$ the total decay width increases and,
correspondingly, the decay path decreases due to the tau Yukawa
coupling and the bottom Yukawa coupling.

\subsection{Neutralino Branching Ratios}

As discussed in the beginning of this section, the lightest neutralino
$\chiz{1}$ will typically decay in the detector. In the following we
present our results for the branching ratios of all R-parity violating
3-body decay of $\chiz{1}$, and of the radiative decay $\chiz{1} \to
\nu_3 \gamma$.  The Feynman diagrams for the decays $\chiz{1} \to
\nu_3 f \bar{f}$ ($f=e$, $\nu_e$, $\mu$, $\nu_\mu$, $u$, $d$, $c$,
$s$, $b$) are shown in \fig{fig:graphs}. For this class of decays we
have $Z^0$, $P^0_i$, and $S^0_j$ exchange in the direct channel
(\fig{fig:graphs}a and b) and $\tilde f$ exchange in the crossed
channels (\fig{fig:graphs}c and d). In particular in the case $f=b$ the
$P^0_i$ and $S^0_j$ exchange contributions are significant.  This is
quite analogous to the results found in \cite{Neut99} for $\chiz{2}
\to \chiz{1} f \bar{f}$ decays.  The particles exchanged in the $s$-,
$t$-, and $u$-channel for the decays $\chiz{1} \to \tau^\pm l^\mp
\nu_l$, ($l=e,\mu$), $\chiz{1} \to \tau^\pm q \bar{q}'$
($q,q'=u,d,s,c$), $\chiz{1} \to \tau^- \tau^+ \nu_l$, and $\chiz{1}
\to 3 \nu_3$ are given in \tab{tab:contribution}.

In the calculations we have included all mixing effects, in particular
the standard MSSM ${\tilde f}_L - {\tilde f}_R$ mixing effects and
those induced by the bilinear R-parity violating terms, i.e.
$Re(\tilde \nu_\tau) - h^0 - H^0$, 
$Im(\tilde \nu_3) - A^0 - G^0$, 
\cite{deCampos:1995av}, ${\tilde \tau}_{L,R}^\pm - H^\pm - G^\pm$
\cite{Akeroyd:1998iq}, $\nu_\tau$ - $\chiz{i}$ \cite{mnutreeJ}, and
$\tau$ - $\chim{j}$ mixings \cite{chitau}. These mixing effects are
particularly important in the calculations of the various R-parity
violating decay rates of $\chiz{1}$, which are discussed below.

\begin{figure}
\setlength{\unitlength}{1mm}
\begin{picture}(150,100)
\put(-3,0){\mbox{\epsfig{figure=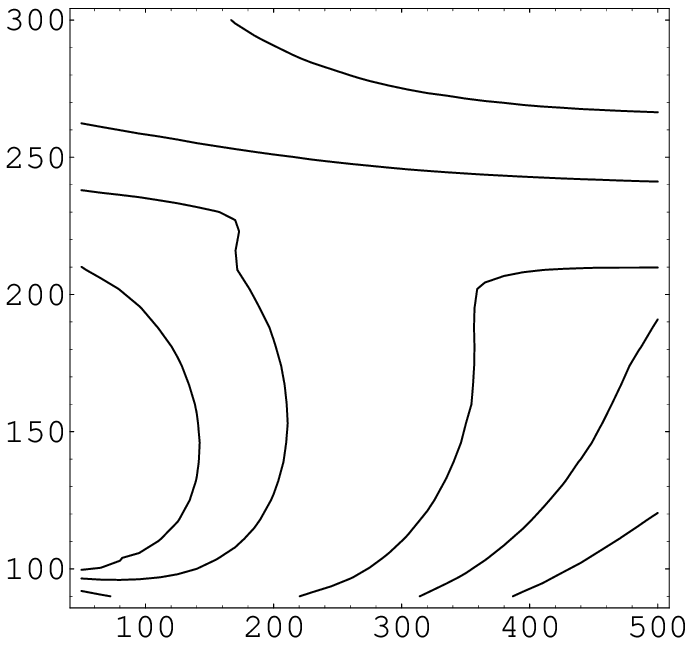,height=8.7cm,width=7.cm}}}
\put(-1,89){{\small \bf a)}}
\put(5,87){\makebox(0,0)[bl]{{\small $M_{1/2}$~[GeV]}}}
\put(39,74){\makebox(0,0)[bl]{{\small $5$}}}
\put(26,66){\makebox(0,0)[bl]{{\small $3$}}}
\put(58,12){\makebox(0,0)[bl]{{\small $7$}}}
\put(57,23){\makebox(0,0)[bl]{{\small $5$}}}
\put(47,42){\makebox(0,0)[bl]{{\small $3$}}}
\put(26,42){\makebox(0,0)[bl]{{\small $1$}}}
\put(18,25){\makebox(0,0)[bl]{{\small $0.5$}}}
\put(7,75){\makebox(0,0)[bl]{{\small $\tan\beta =3$}}}
\put(69,-3){\makebox(0,0)[br]{{ $m_0$~[GeV]}}}
\put(73,0){\mbox{\epsfig{figure=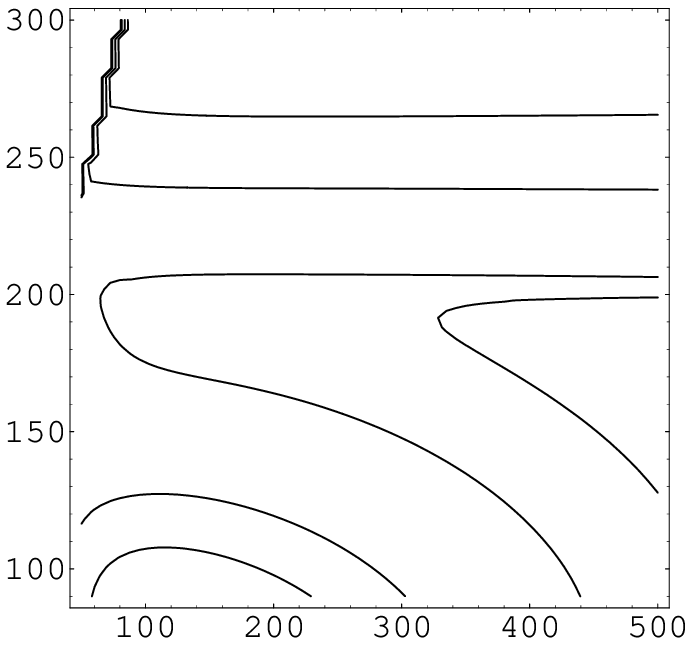,height=8.7cm,width=7.cm}}}
\put(75,89){{\small \bf b)}}
\put(81,87){\makebox(0,0)[bl]{{\small $M_{1/2}$~[GeV]}}}
\put(105,75){\makebox(0,0)[bl]{{\small $\tan\beta =50$}}}
\put(94,71){\makebox(0,0)[bl]{{\small $5$}}}
\put(94,62.5){\makebox(0,0)[bl]{{\small $3$}}}
\put(117,40){\makebox(0,0)[bl]{{\small $5$}}}
\put(94,51){\makebox(0,0)[bl]{{\small $3$}}}
\put(92,23){\makebox(0,0)[bl]{{\small $1$}}}
\put(92,15){\makebox(0,0)[bl]{{\small $0.5$}}}
\put(143,-3){\makebox(0,0)[br]{{ $m_0$~[GeV]}}}
\end{picture}
\caption[]{Branching ratios for  $\chiz{1} \to 3 \, \nu$ in \%
in the $m_0$--$M_{1/2}$ plane for a) $\tan\beta=3$, and b)  $\tan\beta=50$.
 The R-parity violating parameters
are fixed such that $m_{\nu_3} = 0.06$~GeV.}
\label{fig:Bneut3Sugra}
\end{figure}

\begin{figure}
\setlength{\unitlength}{1mm}
\begin{picture}(150,100)
\put(-3,0){\mbox{\epsfig{figure=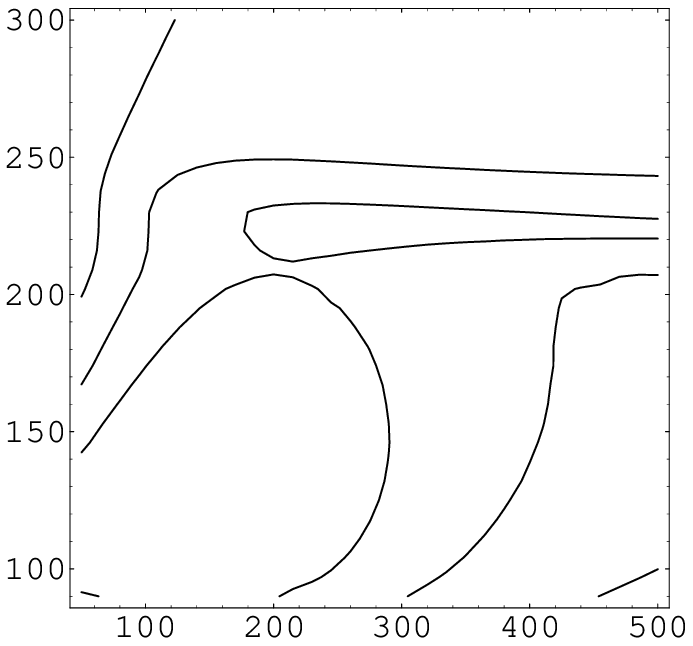,height=8.7cm,width=7.cm}}}
\put(-1,89){{\small \bf a)}}
\put(5,87){\makebox(0,0)[bl]{{\small $M_{1/2}$~[GeV]}}}
\put(14,76){\makebox(0,0)[bl]{{\small $2$}}}
\put(21,66){\makebox(0,0)[bl]{{\small $1$}}}
\put(27,54){\makebox(0,0)[bl]{{\small $0.5$}}}
\put(56,10){\makebox(0,0)[bl]{{\small $2$}}}
\put(50,41){\makebox(0,0)[bl]{{\small $1$}}}
\put(29.5,26){\makebox(0,0)[bl]{{\small $0.5$}}}
\put(25,75){\makebox(0,0)[bl]{{\small $\tan\beta =3$}}}
\put(69,-3){\makebox(0,0)[br]{{ $m_0$~[GeV]}}}
\put(73,0){\mbox{\epsfig{figure=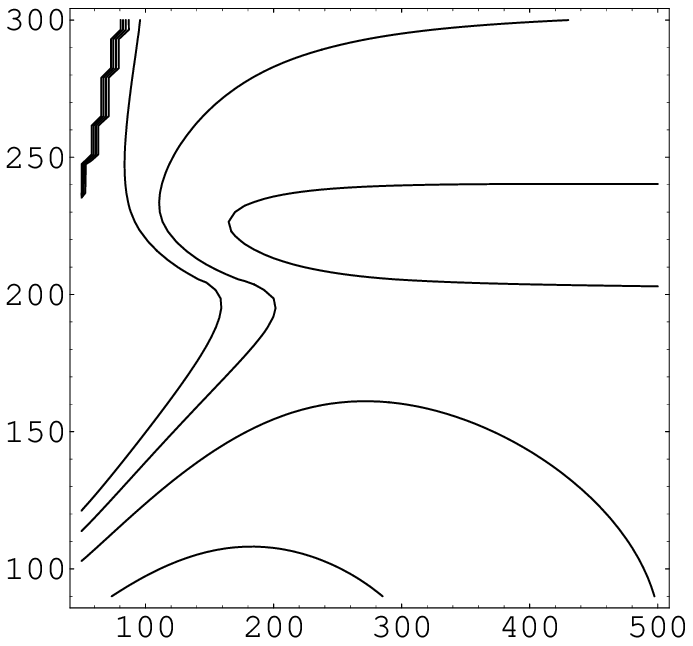,height=8.7cm,width=7.cm}}}
\put(75,89){{\small \bf b)}}
\put(81,87){\makebox(0,0)[bl]{{\small $M_{1/2}$~[GeV]}}}
\put(107,40){\makebox(0,0)[bl]{{\small $\tan\beta =50$}}}
\put(83,62){\makebox(0,0)[bl]{{\small $3$}}}
\put(100,78){\makebox(0,0)[bl]{{\small $2$}}}
\put(100,61){\makebox(0,0)[bl]{{\small $1$}}}
\put(100,33.5){\makebox(0,0)[bl]{{\small $1$}}}
\put(100,16){\makebox(0,0)[bl]{{\small $0.2$}}}
\put(143,-3){\makebox(0,0)[br]{{ $m_0$~[GeV]}}}
\end{picture}
\caption[]{Branching ratios for $\chiz{1} \to \nu_3 \, l^+ \, l^-$ 
in \% in the $m_0$--$M_{1/2}$ plane for a) $\tan\beta=3$,
 and b)  $\tan\beta=50$. Here $l$ is the sum of $e$ and $\mu$.
 The R-parity violating parameters
are fixed such that $m_{\nu_3} = 0.06$~GeV.}
\label{fig:BnllSugra}
\end{figure}

\begin{figure}
\setlength{\unitlength}{1mm}
\begin{picture}(150,100)
\put(-3,0){\mbox{\epsfig{figure=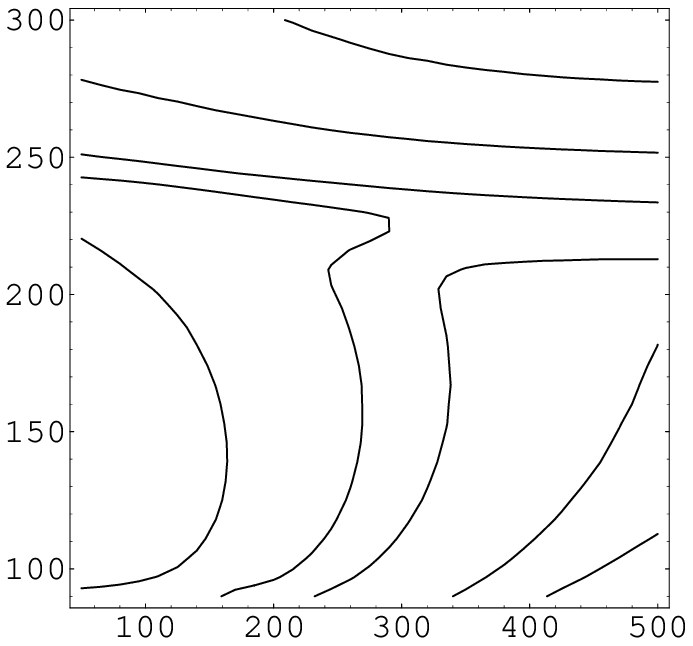,height=8.7cm,width=7.cm}}}
\put(-1,89){{\small \bf a)}}
\put(5,87){\makebox(0,0)[bl]{{\small $M_{1/2}$~[GeV]}}}
\put(39,79){\makebox(0,0)[bl]{{\small $15$}}}
\put(39,68){\makebox(0,0)[bl]{{\small $10$}}}
\put(26,64){\makebox(0,0)[bl]{{\small $5$}}}
\put(57,16){\makebox(0,0)[bl]{{\small $15$}}}
\put(57,22){\makebox(0,0)[bl]{{\small $10$}}}
\put(45,42){\makebox(0,0)[bl]{{\small $5$}}}
\put(34,42){\makebox(0,0)[bl]{{\small $3$}}}
\put(21,25){\makebox(0,0)[bl]{{\small $1$}}}
\put(7,76){\makebox(0,0)[bl]{{\small $\tan\beta =3$}}}
\put(69,-3){\makebox(0,0)[br]{{ $m_0$~[GeV]}}}
\put(73,0){\mbox{\epsfig{figure=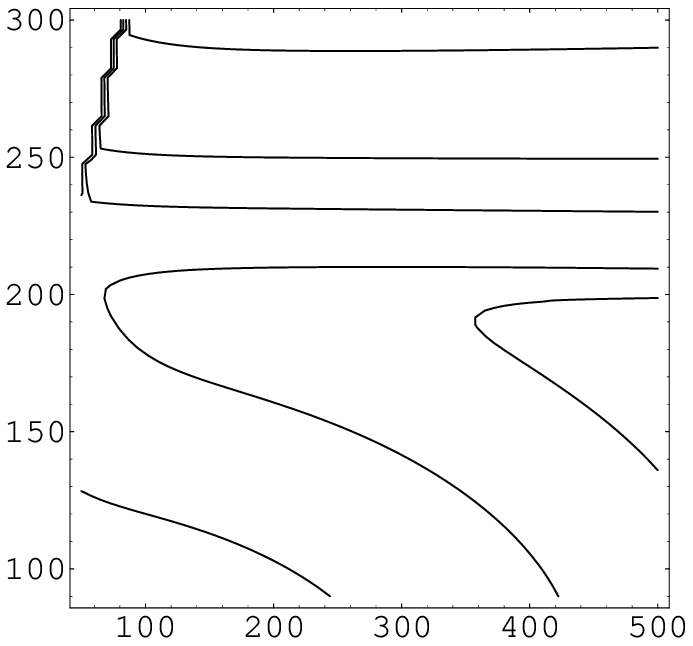,height=8.7cm,width=7.cm}}}
\put(75,89){{\small \bf b)}}
\put(81,87){\makebox(0,0)[bl]{{\small $M_{1/2}$~[GeV]}}}
\put(100,40){\makebox(0,0)[bl]{{\small $\tan\beta =50$}}}
\put(94,79.5){\makebox(0,0)[bl]{{\small $15$}}}
\put(94,66){\makebox(0,0)[bl]{{\small $10$}}}
\put(94,59){\makebox(0,0)[bl]{{\small $5$}}}
\put(94,52){\makebox(0,0)[bl]{{\small $5$}}}
\put(128,40){\makebox(0,0)[bl]{{\small $10$}}}
\put(92,19){\makebox(0,0)[bl]{{\small $1$}}}
\put(143,-3){\makebox(0,0)[br]{{ $m_0$~[GeV]}}}
\end{picture}
\caption[]{Branching ratios for $\chiz{1} \to \nu_3 \, q \, \bar{q}$ in \%
in the $m_0$--$M_{1/2}$ plane for a) $\tan\beta=3$, and b)  $\tan\beta=50$.
Here $q$ is the sum over $u$, $d$, $s$, and $c$.
 The R-parity violating parameters
are fixed such that $m_{\nu_3} = 0.06$~GeV.}
\label{fig:BnqqSugra}
\end{figure}

\begin{figure}
\setlength{\unitlength}{1mm}
\begin{picture}(150,100)
\put(-3,0){\mbox{\epsfig{figure=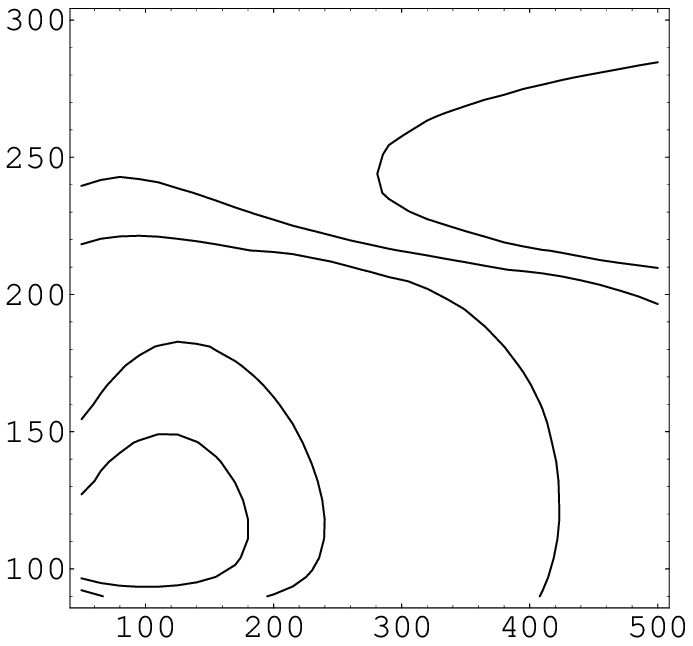,height=8.7cm,width=7.cm}}}
\put(-1,89){{\small \bf a)}}
\put(5,87){\makebox(0,0)[bl]{{\small $M_{1/2}$~[GeV]}}}
\put(39,65){\makebox(0,0)[bl]{{\small $15$}}}
\put(18,63){\makebox(0,0)[bl]{{\small $10$}}}
\put(43,42){\makebox(0,0)[bl]{{\small $5$}}}
\put(19,36){\makebox(0,0)[bl]{{\small $1$}}}
\put(7,20){\makebox(0,0)[bl]{{\small $0.1$}}}
\put(7,75){\makebox(0,0)[bl]{{\small $\tan\beta =3$}}}
\put(69,-3){\makebox(0,0)[br]{{ $m_0$~[GeV]}}}
\put(73,0){\mbox{\epsfig{figure=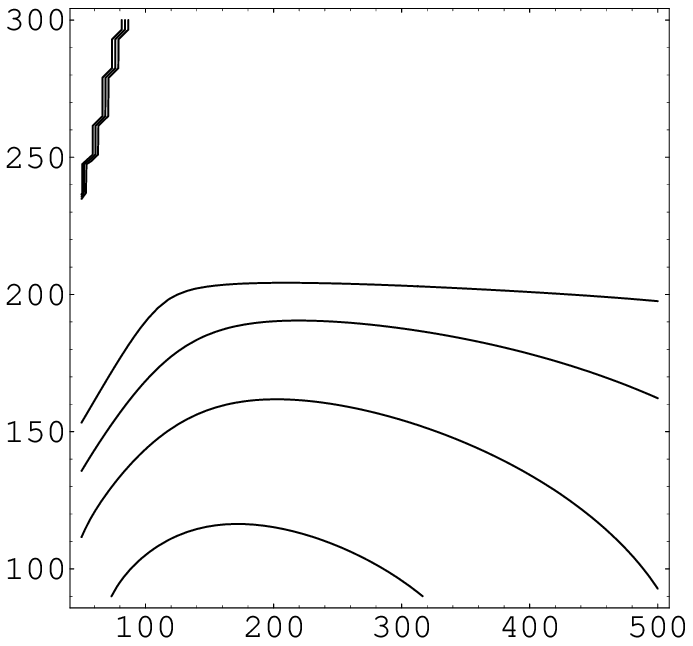,height=8.7cm,width=7.cm}}}
\put(75,89){{\small \bf b)}}
\put(81,87){\makebox(0,0)[bl]{{\small $M_{1/2}$~[GeV]}}}
\put(103,75){\makebox(0,0)[bl]{{\small $\tan\beta =50$}}}
\put(86,77){\vector(-1,0){4}}
\put(86,77){\makebox(0,0)[bl]{{\small $2.6$}}}
\put(85,70){\makebox(0,0)[bl]{{\small $15$}}}
\put(95,50){\makebox(0,0)[bl]{{\small $15$}}}
\put(95,45){\makebox(0,0)[bl]{{\small $10$}}}
\put(95,36){\makebox(0,0)[bl]{{\small $5$}}}
\put(95,20){\makebox(0,0)[bl]{{\small $1$}}}
\put(143,-3){\makebox(0,0)[br]{{ $m_0$~[GeV]}}}
\end{picture}
\caption[]{Branching ratios for $\chiz{1} \to \nu_l \, \tau^{\pm} \, l^{\mp}$ 
in \% in the $m_0$--$M_{1/2}$ plane for a) $\tan\beta=3$,
 and b)  $\tan\beta=50$. Here $l$ is the sum of $e$ and $\mu$.
 The R-parity violating parameters
are fixed such that $m_{\nu_3} = 0.06$~GeV.}
\label{fig:BnetSugra}
\end{figure}

\begin{figure}
\setlength{\unitlength}{1mm}
\begin{picture}(150,100)
\put(-3,0){\mbox{\epsfig{figure=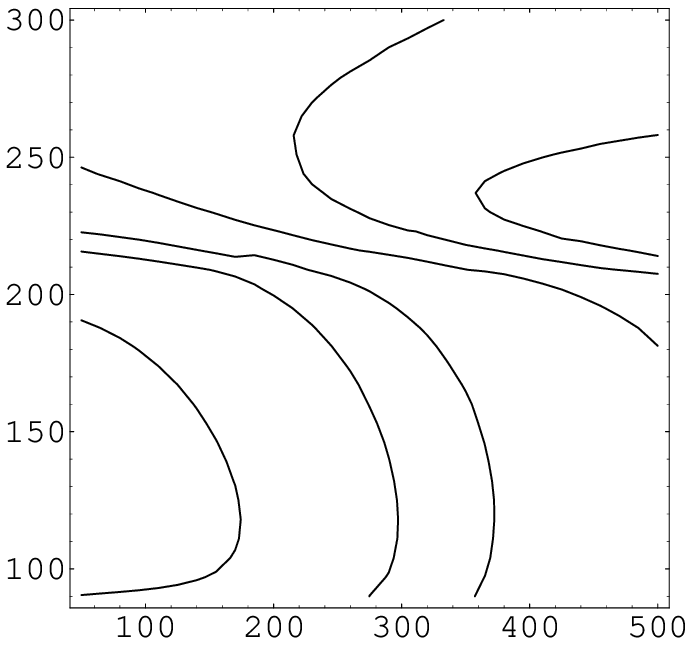,height=8.7cm,width=7.cm}}}
\put(-1,89){{\small \bf a)}}
\put(5,87){\makebox(0,0)[bl]{{\small $M_{1/2}$~[GeV]}}}
\put(42,63){\makebox(0,0)[bl]{{\small $50$}}}
\put(22,63){\makebox(0,0)[bl]{{\small $40$}}}
\put(55,42){\makebox(0,0)[bl]{{\small $25$}}}
\put(34,42){\makebox(0,0)[bl]{{\small $10$}}}
\put(24,42){\makebox(0,0)[bl]{{\small $5$}}}
\put(16,25){\makebox(0,0)[bl]{{\small $1$}}}
\put(12,75){\makebox(0,0)[bl]{{\small $\tan\beta =3$}}}
\put(69,-3){\makebox(0,0)[br]{{ $m_0$~[GeV]}}}
\put(73,0){\mbox{\epsfig{figure=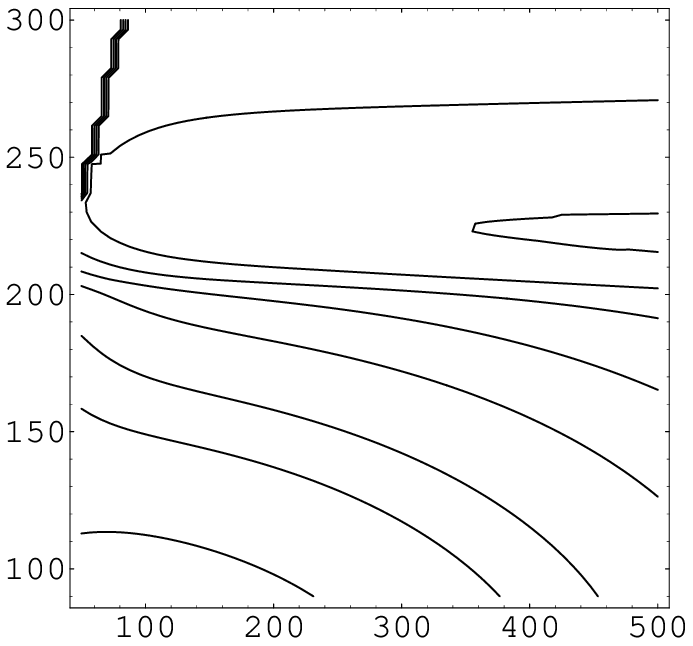,height=8.7cm,width=7.cm}}}
\put(75,89){{\small \bf b)}}
\put(81,87){\makebox(0,0)[bl]{{\small $M_{1/2}$~[GeV]}}}
\put(103,65){\makebox(0,0)[bl]{{\small $\tan\beta =50$}}}
\put(82,77){\makebox(0,0)[bl]{{\small $7$}}}
\put(85,72){\makebox(0,0)[bl]{{\small $50$}}}
\put(100,51.5){\makebox(0,0)[bl]{{\small $50$}}}
\put(130,59){\makebox(0,0)[bl]{{\small $60$}}}
\put(129,46){\makebox(0,0)[bl]{{\small $40$}}}
\put(129,41){\makebox(0,0)[bl]{{\small $30$}}}
\put(100,43){\makebox(0,0)[bl]{{\small $20$}}}
\put(100,34){\makebox(0,0)[bl]{{\small $10$}}}
\put(100,27){\makebox(0,0)[bl]{{\small $5$}}}
\put(100,14){\makebox(0,0)[bl]{{\small $1$}}}
\put(143,-3){\makebox(0,0)[br]{{ $m_0$~[GeV]}}}
\end{picture}
\caption[]{Branching ratios for $\chiz{1} \to \tau^{\pm} \, q \, \bar{q}'$ 
in \% in the $m_0$--$M_{1/2}$ plane for a) $\tan\beta=3$,
 and b)  $\tan\beta=50$.
Here $q$ is the sum over $u$, $d$, $s$, and $c$.
 The R-parity violating parameters
are fixed such that $m_{\nu_3} = 0.06$~GeV.}
\label{fig:BtqqpSugra}
\end{figure}

\begin{figure}
\setlength{\unitlength}{1mm}
\begin{picture}(150,100)
\put(-3,0){\mbox{\epsfig{figure=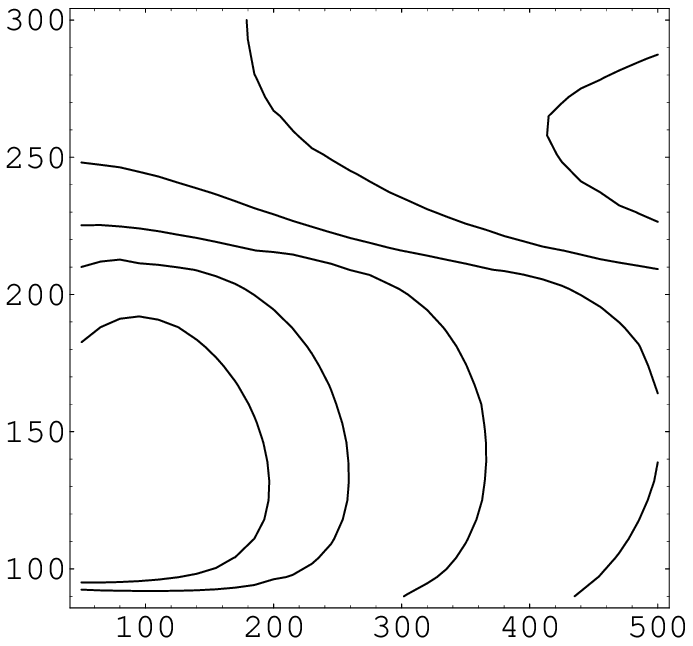,height=8.7cm,width=7.cm}}}
\put(-1,89){{\small \bf a)}}
\put(5,87){\makebox(0,0)[bl]{{\small $M_{1/2}$~[GeV]}}}
\put(48,67){\makebox(0,0)[bl]{{\small $10$}}}
\put(26,62){\makebox(0,0)[bl]{{\small $25$}}}
\put(55,42){\makebox(0,0)[bl]{{\small $50$}}}
\put(36,42){\makebox(0,0)[bl]{{\small $75$}}}
\put(21,42){\makebox(0,0)[bl]{{\small $90$}}}
\put(19,25){\makebox(0,0)[bl]{{\small $95$}}}
\put(27,75){\makebox(0,0)[bl]{{\small $\tan\beta =3$}}}
\put(69,-3){\makebox(0,0)[br]{{ $m_0$~[GeV]}}}
\put(73,0){\mbox{\epsfig{figure=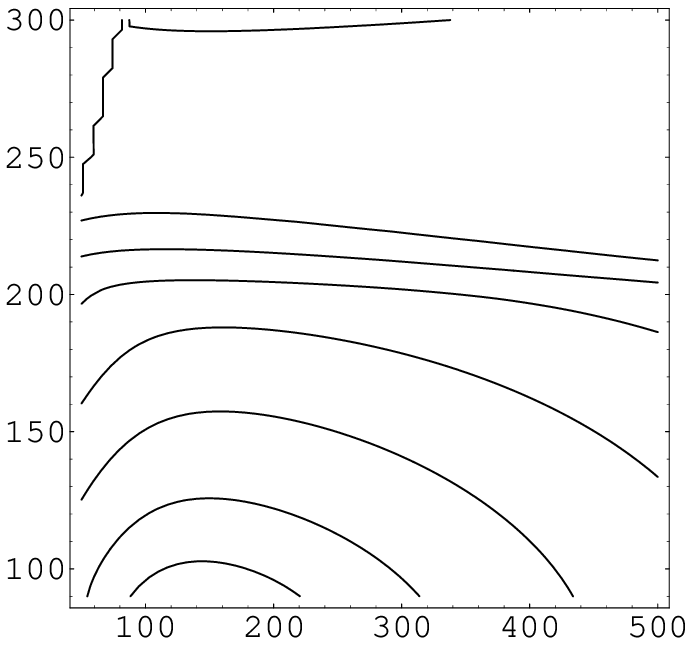,height=8.7cm,width=7.cm}}}
\put(75,89){{\small \bf b)}}
\put(81,87){\makebox(0,0)[bl]{{\small $M_{1/2}$~[GeV]}}}
\put(110,72){\makebox(0,0)[bl]{{\small $\tan\beta =50$}}}
\put(84,72){\makebox(0,0)[bl]{{\small $1$}}}
\put(92,77){\makebox(0,0)[bl]{{\small $5$}}}
\put(94,59){\makebox(0,0)[bl]{{\small $5$}}}
\put(92,54.5){\makebox(0,0)[bl]{{\small $10$}}}
\put(92,50){\makebox(0,0)[bl]{{\small $25$}}}
\put(92,45){\makebox(0,0)[bl]{{\small $50$}}}
\put(92,34){\makebox(0,0)[bl]{{\small $75$}}}
\put(92,23){\makebox(0,0)[bl]{{\small $90$}}}
\put(92,15){\makebox(0,0)[bl]{{\small $95$}}}
\put(143,-3){\makebox(0,0)[br]{{ $m_0$~[GeV]}}}
\end{picture}
\caption[]{Branching ratios for $\chiz{1} \to \nu_3 \, b \, \bar{b}$ in \%
in the $m_0$--$M_{1/2}$ plane for a) $\tan\beta=3$, and b)  $\tan\beta=50$.
 The R-parity violating parameters
are fixed such that $m_{\nu_3} = 0.06$~GeV.}
\label{fig:BnbbSugra}
\end{figure}

\begin{figure}
\setlength{\unitlength}{1mm}
\begin{picture}(150,100)
\put(-3,0){\mbox{\epsfig{figure=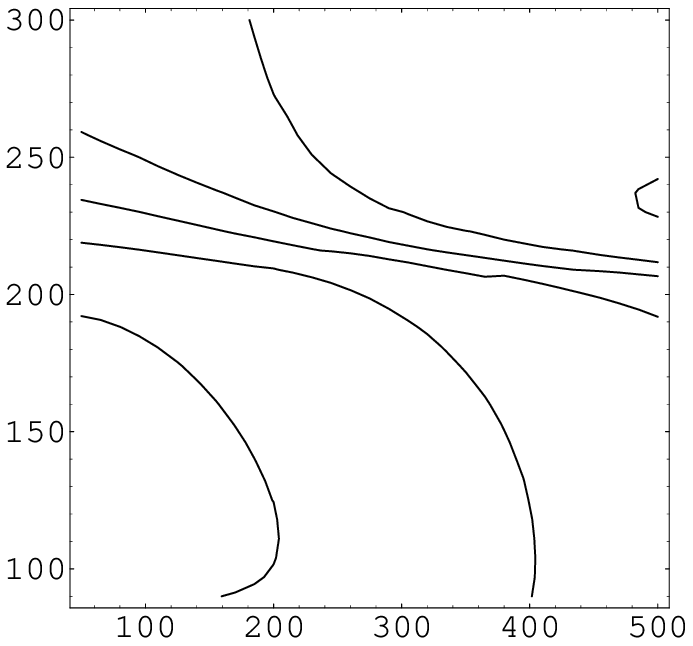,height=8.7cm,width=7.cm}}}
\put(-1,89){{\small \bf a)}}
\put(5,87){\makebox(0,0)[bl]{{\small $M_{1/2}$~[GeV]}}}
\put(59,62){\makebox(0,0)[bl]{{\small $9$}}}
\put(32,63){\makebox(0,0)[bl]{{\small $7$}}}
\put(21,62){\makebox(0,0)[bl]{{\small $5$}}}
\put(54,44){\makebox(0,0)[bl]{{\small $3$}}}
\put(34,42){\makebox(0,0)[bl]{{\small $1$}}}
\put(24,26){\makebox(0,0)[bl]{{\small $0.1$}}}
\put(35,75){\makebox(0,0)[bl]{{\small $\tan\beta =3$}}}
\put(69,-3){\makebox(0,0)[br]{{ $m_0$~[GeV]}}}
\put(73,0){\mbox{\epsfig{figure=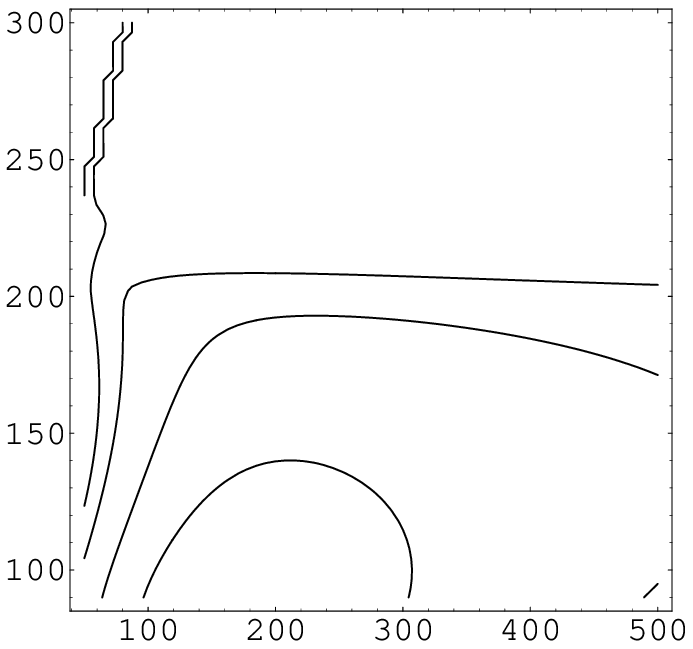,height=8.7cm,width=7.cm}}}
\put(75,89){{\small \bf b)}}
\put(81,87){\makebox(0,0)[bl]{{\small $M_{1/2}$~[GeV]}}}
\put(105,72){\makebox(0,0)[bl]{{\small $\tan\beta =50$}}}
\put(87,77){\vector(-1,0){5}}
\put(87,77){\makebox(0,0)[bl]{{\small $90.3$}}}
\put(85,69){\makebox(0,0)[bl]{{\small $10$}}}
\put(100,52){\makebox(0,0)[bl]{{\small $8$}}}
\put(100,46){\makebox(0,0)[bl]{{\small $5$}}}
\put(100,28){\makebox(0,0)[bl]{{\small $3$}}}
\put(143,-3){\makebox(0,0)[br]{{ $m_0$~[GeV]}}}
\end{picture}
\caption[]{Branching ratios for $\chiz{1} \to \nu_3 \, \tau^+ \, \tau^-$ 
in \% in the $m_0$--$M_{1/2}$ plane for a) $\tan\beta=3$,
 and b)  $\tan\beta=50$.  The R-parity violating parameters
are fixed such that $m_{\nu_3} = 0.06$~GeV.}
\label{fig:BnttSugra}
\end{figure}

\begin{figure}
\setlength{\unitlength}{1mm}
\begin{picture}(150,100)
\put(-3,0){\mbox{\epsfig{figure=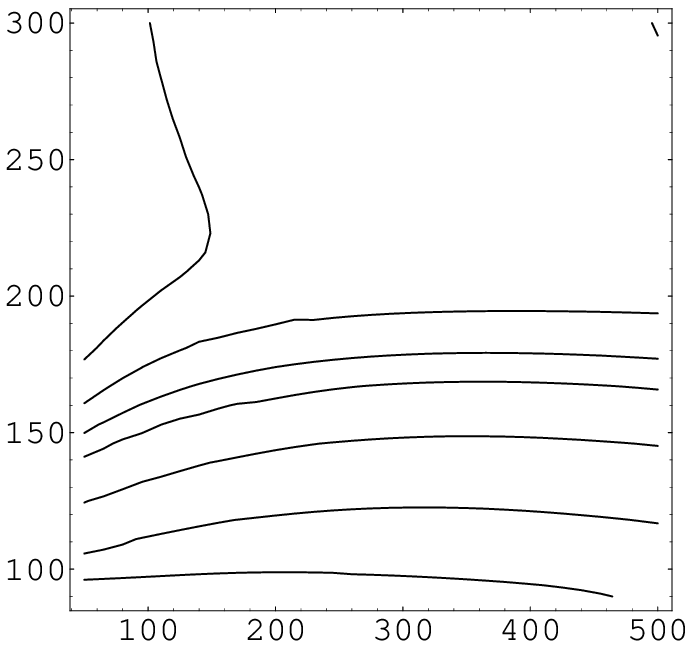,height=8.7cm,width=7.cm}}}
\put(-1,89){{\small \bf a)}}
\put(5,87){\makebox(0,0)[bl]{{\small $M_{1/2}$~[GeV]}}}
\put(18,61.5){\makebox(0,0)[bl]{{\small $10^{-3}$}}}
\put(30,46){\makebox(0,0)[bl]{{\small $10^{-4}$}}}
\put(30,41){\makebox(0,0)[bl]{{\small $10^{-4}$}}}
\put(21,35.5){\makebox(0,0)[bl]{{\small $10^{-3}$}}}
\put(30,31){\makebox(0,0)[bl]{{\small $0.01$}}}
\put(30,22){\makebox(0,0)[bl]{{\small $0.1$}}}
\put(30,14){\makebox(0,0)[bl]{{\small $1$}}}
\put(30,75){\makebox(0,0)[bl]{{\small $\tan\beta =3$}}}
\put(69,-3){\makebox(0,0)[br]{{ $m_0$~[GeV]}}}
\put(73,0){\mbox{\epsfig{figure=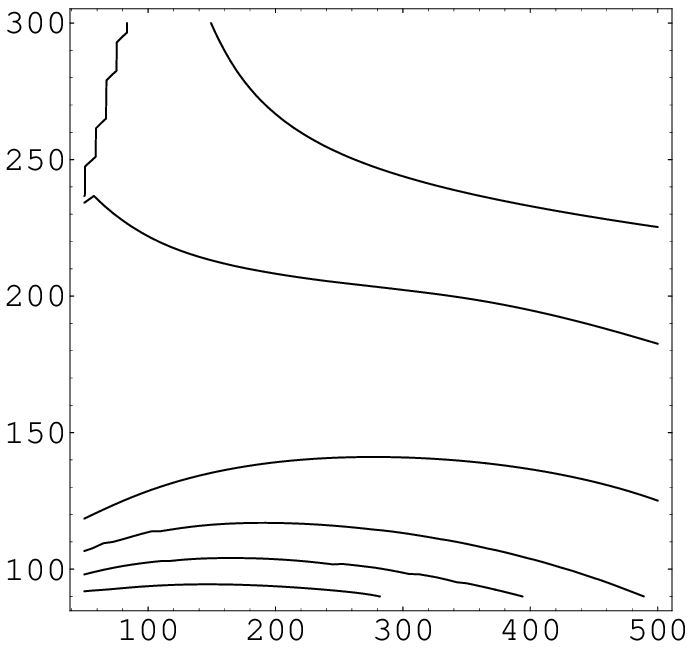,height=8.7cm,width=7.cm}}}
\put(75,89){{\small \bf b)}}
\put(81,87){\makebox(0,0)[bl]{{\small $M_{1/2}$~[GeV]}}}
\put(87,35){\makebox(0,0)[bl]{{\small $\tan\beta =50$}}}
\put(85,72){\makebox(0,0)[bl]{{\small $10^{-3}$}}}
\put(105,68){\makebox(0,0)[bl]{{\small $10^{-2}$}}}
\put(105,51){\makebox(0,0)[bl]{{\small $10^{-3}$}}}
\put(105,28){\makebox(0,0)[bl]{{\small $10^{-3}$}}}
\put(105,14.5){\makebox(0,0)[bl]{{\small $10^{-2}$}}}
\put(105,10){\makebox(0,0)[bl]{{\small $10^{-3}$}}}
\put(143,-3){\makebox(0,0)[br]{{ $m_0$~[GeV]}}}
\end{picture}
\caption[]{Branching ratios for $\chiz{1} \to \nu_3 \, \gamma$ 
in \% in the $m_0$--$M_{1/2}$ plane for a) $\tan\beta=3$,
 and b)  $\tan\beta=50$.  The R-parity violating parameters
are fixed such that $m_{\nu_3} = 0.06$~GeV.}
\label{fig:BngSugra}
\end{figure}

In the following plots \fig{fig:Bneut3Sugra} - \ref{fig:BngSugra} we
show contour lines in the $m_0$-$M_{1/2}$ plane for the branching
ratios in \% of the various $\chiz{1}$ decays, in (a) for $\tanb = 3$
and in (b) for $\tanb=50$. We have fixed the mass of the heaviest
neutrino to $m_{\nu_3} = 0.06 $~eV \cite{atm99}. It turns out, that in
the range $10^{-2}$~eV $\leq m_{\nu_3} \leq 1$~keV all the $\chiz{1}$
decay branching ratios are rather insensitive to the actual value of
$m_{\nu_3}$.  This is an important feature of our supergravity--type
R-parity violating model. It is a consequence of the fact that, as a
result of the universal supergravity boundary conditions on the soft
breaking terms, all R-parity violating couplings are proportional to a
unique common parameter which may be taken as $\epsilon_3 / \mu$. For
a more detailed discussion on this proportionality the reader is
referred to ref.~\cite{nulong5}.
Also note that for $M_{1/2} \gsim$ 220~GeV the neutralino mass becomes
larger than $m_W$ and $m_Z$ so that $\chiz{1}$ decays into real $W$
and $Z$ are possible. The effects of these real decays can be seen for
$M_{1/2} \gsim$ 220~GeV in most of the following plots.
For the large $\tan\beta$ case ($\tan\beta = 50$) and $M_{1/2} \gg M_0$
the mass of the lighter charged boson $S^\pm_1$ is smaller than
$\mchiz{1}$ (upper left corner of \fig{fig:Bneut3Sugra}b -
\ref{fig:BngSugra}b). In this region of the parameter space the two
2-body decays $\chiz{1} \to W^\pm \tau_\pm$ and $\chiz{1} \to S^\pm_1
\tau_\pm$ compete.  The first one is R-parity violating, but has more
phase space than the second one which is R-parity conserving, since
$S^\pm_1$ is mainly a stau. For this reason, the most import final
state is $\tau^+ \tau^- \nu_3$, followed by $\tau^\pm q \bar{q}'$ and
$\tau^\pm l^\mp \nu_i$ ($l=e,\mu$) as shown in
Figs.~\ref{fig:BnttSugra}, \ref{fig:BtqqpSugra}, and
\ref{fig:BnetSugra}, respectively.  All other final states have nearly
vanishing branching ratios in this corner of the parameter space.

\fig{fig:Bneut3Sugra}a and b exhibit the contour lines for the
branching ratio of the invisible decay $\chiz{1} \to 3 \, \nu$. This
branching ratio can reach 7\% for the parameters chosen. In
\figs{fig:BnllSugra}{fig:BnqqSugra} we show the branching
ratio for the decays $\chiz{1} \to \nu_3 \, l^+ \, l^-$ and
$\chiz{1} \to \nu_3 \, q \, \bar{q}$ where $l$ and $q$ denote the
leptons and quarks of the first two generations, summed over all
flavors. These branching ratios can go up to 3\% and 15\%,
respectively. Notice that the sneutrino, slepton, and squark exchange
contributions to the $\chiz{1}$ decays become larger with increasing
$m_0$, despite the fact that the increase of the scalar masses
$m_{\tilde \nu}$, $m_{\tilde l}$, $m_{\tilde q}$ suppresses these
exchange contributions. 
This trend can also be observed in \fig{fig:BnqqSugra},
\ref{fig:BnetSugra} and \ref{fig:BtqqpSugra}.
This happens because the tadpole equations correlate $\mu$ to $m_0$.
Increasing $\mu$ while keeping $M_1$ and $M_2$ fixed implies
increasing the gaugino content of $\chiz{1}$ and, hence, enhancing the
$\chiz{1}$-$f$-$\tilde f$ couplings.

In \figs{fig:BnetSugra}{fig:BtqqpSugra} we show the contour
lines for the branching ratios of the LSP decays involving a single
tau, namely $\chiz{1} \to \nu_l \, \tau^{\pm} \, l^{\mp}$ and
$\chiz{1} \to \tau^{\pm} \, q \, \bar{q}'$, where $l$, $q$, and $q'$
are summed over the first two generations.  The branching for these
decay modes can reach up to 20\% and 60\% respectively. For
$M_{1/2}\gsim 220$~GeV decays into real $W^\pm$ dominate. If this is
the case and if both $\chiz{1}$ produced in $e^+ e^- \to \chiz{1}
\chiz{1}$ decay according to these modes this would lead to very
distinctive final states, such as $4 j \tau^+ \tau^+$, $\tau^+ \tau^+
l^- l^-$ ($l=e,\mu$), or $\tau^+ \tau^+ e^- \mu^-$. The full list of
expected signals is given in \tab{tab:signals}.  The first column in this table
specifies the two pairs of $\chiz{1}$ decay modes, while the second
one gives the corresponding signature. In the last column we state
whether the corresponding signature exists for $e^+ e^- \to \chiz{1}
\chiz{2}$ production within mSUGRA.

\begin{table}
\begin{center}
\begin{tabular}{|l|c|c|}
\hline
Decay mode & exchanged particle & channel \\ \hline
$\chiz{1} \to 3 \, \nu_3$ & $Z$, $S^0_i$, $P^0_j$  & $s$ \\
                          & $Z$, $S^0_i$, $P^0_j$  & $t$ \\
                          & $Z$, $S^0_i$, $P^0_j$  & $u$ \\ \hline
$\chiz{1} \to \nu_3 \, \nu_l \, \bar{\nu}_l$ ($l=e,\mu$) & $Z$  & $s$ \\
                          & $\overline{\tilde \nu}_l$  & $t$ \\
                          & $\tilde \nu_l$ & $u$ \\ \hline
$\chiz{1} \to \nu_3 \, f \, \bar{f}$ ($f=e,\mu,u,d,s,c,b$)
                                         & $Z$, $S^0_i$, $P^0_j$ & $s$ \\
                          & $\overline{\tilde f}_{1,2}$  & $t$ \\
                          & $\tilde f_{1,2}$ & $u$ \\ \hline
$\chiz{1} \to \nu_3 \, \tau^+ \, \tau^-$&  $Z$, $S^0_i$, $P^0_j$ & $s$ \\
                          &  $W^-$, $S^-_k$ & $t$ \\
                          &  $W^+$, $S^+_k$ & $u$ \\ \hline
$\chiz{1} \to \nu_l \, \tau^{\pm} \, l^{\mp}$ ($l=e,\mu$) &
                             $W^\pm$, $S^\pm_k$ & $s$ \\
                          & $\overline{\tilde l}_{1,2}$  & $t$ \\
                          & $\tilde \nu_l$ & $u$ \\ \hline
$\chiz{1} \to \tau \, q \, \bar{q}'$ ($q=u,c$, $q'=d,s$) &
                             $W^\pm$, $S^\pm_k$ & $s$ \\
                          & $\overline{\tilde q}'_{1,2}$  & $t$ \\
                          & $\tilde q_{1,2}$ & $u$ \\ \hline
\end{tabular}
\end{center}
\caption{Contributions involved in the lightest neutralino 3-body 
decay modes. The $s$-, $t$-, and  $u$-channels are defined by: 
$s=(p_1 - p_2)^2$, $t=(p_1 - p_3)^2$, and $u=(p_1 - p_4)^2$. 
See also \fig{fig:graphs}.}
\label{tab:contribution}
\end{table}

The LSP decays involving only third generation fermions, namely,
$\chiz{1} \to \nu_3 \, b \, \bar{b}$ and $\chiz{1} \to \nu_3 \, \tau^+
\, \tau^-$ are different from those into the first and second
generation fermion pairs, because the Higgs boson exchanges and the
Yukawa terms play a very important r\^ole.  This can be seen in
\figs{fig:BnbbSugra}{fig:BnttSugra} , where we plot the
contour lines for these decays.  The branching ratio of $\chiz{1} \to
\nu_3 \, b \, \bar{b}$ can reach up to 97\%. The decay rate is large
because the scalar exchange contributions ($S^0_j,P^0_j,\tilde b_k$)
are large for $M_{1/2} \lsim 200$~GeV. Note that this is also the case
for $\tanb =3$, because not only the neutrino-neutralino mixing
proportional to $m_{\nu_3}$ is important but also the
neutrino-higgsino mixing proportional to $\epsilon_3 / \mu$. The
decrease of the branching ratio with increasing $m_0$ is due to the
decrease of the higgsino component of $\chiz{1}$ and the increase of
the Higgs boson masses.  For $M_{1/2} \gsim 200$~GeV the decays into
real $W^+$ and $Z^0$ are possible, reducing the branching ratio of
$\chiz{1} \to \nu_3 \, b \, \bar{b}$.
As shown in \fig{fig:BnttSugra} the branching ratio for $\chiz{1} \to
\nu_3 \, \tau^+ \, \tau^-$ is very small for $\tanb = 3$ and $M_{1/2}
\lsim 200$~GeV.  This is due to the destructive interference between
$Z^0$ contribution and the contributions of the exchanged charged
scalar particles (mainly due to the stau components of $S^\pm_k$).

Finally we have also considered the radiative LSP decay mode $\chiz{1}
\to \nu_3 \, \gamma$ \cite{muk99}.  In \fig{fig:BngSugra} the
branching ratio for this mode is shown. This decay proceeds only at
one-loop level and therefore is in general suppressed compared to the
three-body decay modes.  However, for $M_{1/2} \lsim 125$~GeV and
large $\tanb$ it exceeds 1\%, leading to interesting signatures like
$e^+ e^- \to \chiz{1} \chiz{1} \to \tau^\pm \, \mu^\mp \,\gamma$ +
\missPT.  Due to initial state radiation it can easily happen that a
second photon is observed in the same event.

The complete list of possible signatures stemming from LSP decays in
our bilinear \rp model is shown in \tab{tab:signals}. In this table we
also indicate whether the same signatures could also arise in mSUGRA as
a result of $e^+ e^- \to \chiz{1} \chiz{2}$ followed by the MSSM decay
modes of $\chiz{2}$ if its production is kinematically allowed.  The
final states 4 jets + \missPT, $\tau$ + 2 jets + \missPT, and $\tau$ +
($e$ or $\mu$) + {\missPT} would also occur in mSUGRA via the decay
of $\chiz{2}$ into $\chipm{1}$.  However one expects in general that
these decay modes are suppressed within mSUGRA. 
In contrast in the R-parity violating case these signatures can be
rather large as can be seen from 
\figs{fig:BnetSugra}{fig:BtqqpSugra}.  
Note moreover, that some of the \rp signatures
are practically background free.  For example, due to the Majorana nature
of $\chiz{1}$, one can have two same--sign $\tau$ leptons + 4 jets +
{\missPT}.  Other interesting signals are: $\tau$ + 3 ($e$ and/or
$\mu$) + \missPT, 3 $\tau$ + ($e$ or $\mu$) + \missPT, $\tau$ + ($e$
or $\mu$) + 2 jets + \missPT, $\tau$ + 4 jets + \missPT, $\tau^\pm
\tau^\pm$ + ($e$ or $\mu$) + 2 jets + \missPT, or $\tau^\pm \tau^\pm$
+ $l^\mp {l'}^\mp$ + {\missPT} with $l=e,\mu$.
In Table~\ref{tab:detail} we give masses and branching ratios for
typical examples.

\begin{table}
\begin{center}
\begin{tabular}{|l|l|l|}
\hline
Combination of $\chiz{1}$ decay modes & signature & mSUGRA-like \\ \hline
$(3 \, \nu)$ $(3 \, \nu)$ & \missPT  & yes \\ \hline
$(3 \, \nu)$ $(\nu_3 l^+ l^-)$ & 2 leptons + \missPT  & yes \\ \hline
$(3 \, \nu)$ $(\nu_3 q \bar{q})$ & 2 jets + \missPT  & yes \\
$(3 \, \nu)$ $(\nu_3 b \bar{b})$ &                   &     \\ \hline
$(3 \, \nu)$ $(\nu_l \tau^\pm l^\mp)$ with $l=e,\mu$ &
      $\tau$ + ($e$ or $\mu$) + \missPT  & yes, but suppressed \\ \hline
$(3 \, \nu)$ $(\tau^\pm q \bar{q}')$ &
      $\tau$ + 2 jets + \missPT  & yes, but suppressed \\ \hline
$(3 \, \nu)$ $(\nu_3 \, \gamma)$ & $\gamma$ + \missPT  & yes \\ \hline
$(\nu_3 l^+ l^-)$ $(\nu_3 {l'}^+ {l'}^-)$ & 4 leptons + \missPT  & no \\ \hline
$(\nu_3 l^+ l^-)$ $(\nu_3 q \bar{q})$ & 2 leptons + 2jets + \missPT 
                                                                 & no \\
$(\nu_3 l^+ l^-)$ $(\nu_3 b \bar{b})$ & & \\ \hline
$(\nu_3 l^+ l^-)$ $(\nu_l \tau^\pm l^\mp)$ with $l=e,\mu$ &
      $\tau$ + 3 ($e$ and/or $\mu$) + \missPT  & no \\ \hline
$(\nu_3 \tau^+ \tau^-)$ $(\nu_l \tau^\pm l^\mp)$ with $l=e,\mu$ &
      3 $\tau$ +  ($e$ or $\mu$) + \missPT  & no \\ \hline
$(\nu_3 l^+ l^-)$ $(\tau^\pm q \bar{q}')$ &
      $\tau$ + 2 leptons + 2 jets + \missPT  & no \\ \hline
$(\nu_3 \tau^+ \tau^-)$ $(\tau^\pm q \bar{q}')$ &
      3 $\tau$  + 2 jets + \missPT  & no \\ \hline
$(\nu_3 l^+ l^-)$ $(\nu_3 \gamma)$ & 2 leptons + $\gamma$ + \missPT
                                                          & no \\ \hline
$(\nu_3 q \bar{q})$ $(\nu_3 q \bar{q})$ & & \\
$(\nu_3 q \bar{q})$ $(\nu_3 b \bar{b})$ & 4 jets + \missPT & yes, but
                                                          suppressed\\
$(\nu_3 b \bar{b})$ $(\nu_3 b \bar{b})$ & & \\ \hline
$(\nu_3 q \bar{q})$ $(\nu_l \tau^\pm l^\mp)$ with $l=e,\mu$ & 
          $\tau$ + ($e$ or $\mu$) + 2 jets + \missPT & no \\
$(\nu_3 b \bar{b})$ $(\nu_l \tau^\pm l^\mp)$ with $l=e,\mu$ & & \\ \hline
$(\nu_3 q \bar{q})$  $(\tau^\pm q \bar{q}')$ & $\tau$ + 4 jets + \missPT
                                                     & no \\
$(\nu_3 b \bar{b})$  $(\tau^\pm q \bar{q}')$ & & \\ \hline
$(\nu_3 q \bar{q})$ $(\nu_3 \gamma)$ & 2 jets + $\gamma$ + \missPT
                                                          & no \\
$(\nu_3 b \bar{b})$ $(\nu_3 \gamma)$ &  &  \\ \hline
$(\nu_l \tau^\pm l^\mp)$ $(\nu_l \tau^\pm {l'}^\mp)$ &
     $\tau^\pm \tau^\pm$ + $l^\mp {l'}^\mp$ + \missPT & no \\
   &  $\tau^\pm \tau^\mp$ + $l^\mp {l'}^\pm$ + \missPT & no \\ \hline
$(\nu_l \tau^\pm l^\mp)$ $(\tau^\pm q \bar{q}')$ &
     $\tau^\pm \tau^\pm$ + ($e$ or $\mu$) + 2 jets + \missPT & no \\
   &  $\tau^\pm \tau^\mp$ + ($e$ or $\mu$) + 2 jets + \missPT & no \\ \hline
$(\nu_l \tau^\pm l^\mp)$  $(\nu_3 \gamma)$ & 
     $\tau$ + ($e$ or $\mu$) + $\gamma$ + \missPT & no \\ \hline
$(\tau^\pm q \bar{q}')$ $(\tau^\pm q \bar{q}')$ &
     $\tau^\pm \tau^\pm$ + 4 jets + \missPT & no \\
   &  $\tau^\pm \tau^\mp$ + 4 jets + \missPT & no \\ \hline
$(\tau^\pm q \bar{q}')$  $(\nu_3 \gamma)$ & 
     $\tau$ + 2 jets + $\gamma$ + \missPT & no \\ \hline
$(\nu_3 \gamma)$ $(\nu_3 \gamma)$ & 2 $\gamma$ + \missPT & no \\ \hline
\end{tabular}
\end{center}
\caption{The signatures  expected from the process 
$e^+ e^- \to \chiz{1} \chiz{1}$ in the bilinear \rp model. }
\label{tab:signals}
\end{table}
\begin{table}[h]
\begin{center}
\begin{tabular}{|c||c|c|c||c|c|c|} \hline
 & \multicolumn{3}{c||}{$\tan \beta = 3$}
 & \multicolumn{3}{c|}{$\tan \beta = 50$} \\ 
 & A & B & C & A & B & C \\ \hline
$\mchiz{1}$ & 54.6 & 59.0 & 92.5 & 60.0 & 61.5 & 94.4 \\ 
$m_{S^0_1}$ & 91.0 & 96.8 & 102.9 & 107.2 & 111.1 & 116.4 \\
$m_{\tilde \nu}$ & 180.5 & 449.6 & 466.0 & 178.2 & 448.7 & 465.1 \\
$m_{\tilde e_R}$ & 170.5 & 445.7 & 450.6 & 171.6 & 446.1 & 451.0 \\
$m_{\tilde e_L}$ & 194.2 & 455.2 & 471.4 & 195.3 & 455.8 & 471.9 \\
$m_{\tilde q}$ & 398.1 & 572.8 & 705.4 &  398.1 & 572.8 & 705.4 \\ 
$m_{\tilde t_1}$ & 261.4 & 328.5 & 442.2 & 279.9 & 355.2 & 466.3 \\
$m_{\tilde b_1}$ & 361.3 & 479.1 & 612.1 & 243.0 & 343.0 & 470.1 \\ \hline
BR$(\chiz{1} \to 3 \nu$)               &  0.5 &  4.5 &  1.8 &  0.3 &  1.2 &
  1.9 \\
BR$(\chiz{1} \to l^- l^+ \nu_3)$       &  0.2 &  1.1 &  0.5 &  0.2 &  0.3 &
  0.6 \\
BR$(\chiz{1} \to q \bar{q} \nu_3)$     &  1.0 &  8.6 &  4.0 &  0.5 &  2.2 &
  4.4 \\
BR$(\chiz{1} \to l^\pm \tau^\mp \nu)$  &  0.6 &  5.6 & 18.0 &  0.5 &  1.8 &
 17.8 \\
BR$(\chiz{1} \to q \bar{q}' \tau^\pm)$ &  1.1 & 16.1 & 53.7 &  0.9 &  5.1 &
 53.2 \\
BR$(\chiz{1} \to b \bar{b} \nu_3)$     & 96.5 & 62.6 & 13.4 & 97.1 & 88.4 &
 13.3 \\
BR$(\chiz{1} \to \tau^- \tau^+ \nu_3)$ &  0.1 &  1.5 &  8.6 &  0.5 &  1.0 &
  8.8 \\
\hline
\end{tabular}
\end{center}
\caption{Masses and branching ratios for the points:
 A $(M_{1/2},m_0)$ = (153,155),
 B  $(M_{1/2},m_0)$ = (153,440), and  C $(M_{1/2},m_0)$ = (251,440) for
both $\tan\beta =3$ and 50.
The masses are given in GeV and the branching ratios in \% and we only
give those larger than 0.1\%. Here the same
summations of the final states are performed as in the figures. 
$m_{\tilde q}$ is the averaged squark mass for the first two generations.}
\label{tab:detail}
\end{table}

As it is well known, also in gauge mediated supersymmetry breaking
models (GMSB) \cite{Giudice:1999bp} the neutralino can decay inside the
detector, because the gravitino $\tilde G$ is the LSP. It is therefore
an interesting question if the R-parity violating model can be
confused with GMSB. To answer this question let us have a look at the
dominant decay modes of the lightest neutralino in GMSB.
If the lightest neutralino is the NLSP, its main decay mode in GMSB is
\begin{eqnarray}
\chiz{1} \to \gamma \, \tilde G \, \, , \nonumber
\end{eqnarray}
where $\tilde G$ is the gravitino.  For the case where at least one of
the sleptons is lighter than the lightest neutralino the latter has
the following decay chain $ \chiz{1} \to {\tilde l}^\pm \, l^\mp \to
l^\pm \, l^\mp \, \tilde G \, \, , $.  In principle three-body decay
modes mediated by virtual photon, virtual Z-boson and virtual
sfermions also exist. However, in the neutralino mass range considered
here these decays are phase--space--supressed
\cite{Giudice:1999bp,Bagger:1997bt}. This implies that the R-parity
violating model can not be confused with GMSB, because (i) in GMSB the
final states containing quarks are strongly suppressed, and (ii) GMSB
with conserved R-parity implies lepton flavour conservation, and
therefore there are no final states like $e^+ e^+ \tau^- \tau^-$ +
\missPT. A further interesting question would be how the neutralino
phenomenology changes in a GMSB scenario with broken R-parity. The
main consequence would be an enhancement of final states containing
photons and/or leptons. A detailed study of this question is, however,
beyond the scope of the present paper.

\section{Conclusions}

We have studied the production of the lightest neutralino $\chiz{1}$
at LEP2 and the resulting phenomenology in models where an effective
bilinear term in the superpotential parametrizes the explicit breaking
of R-parity. We have considered supergravity scenarios which can be
explored at LEP2 in which the lightest neutralino is also the lightest
supersymmetric particle. We have presented a detailed study of the LSP
$\chiz{1}$ decay properties and studied the general features of the
corresponding signals expected at LEP2.  A detailed investigation of
the possible detectability of the signals discussed in
\tab{tab:signals} taking into account realistic detector features is
beyond the scope of this paper.  Clearly, existing LEP2 data are
already probing the part of the parameter region which corresponds to
approximately $\mchiz{1} \lsim 40$~GeV.  Finally, we note that, in
addition to important modifications in the $\chiz{1}$ decay
properties, R-parity violating decay models lead also to new
interesting features in other decays, such as charged
\cite{Akeroyd:1998iq} and neutral \cite{deCampos:1995av} Higgs boson
and slepton decays, stop decays~\cite{stop1,stop2,stop3}, and gluino
cascade decays \cite{gluino}. In addition we have shown that the
R-parity violating model can not be confused with gauge mediated
supersymmetry breaking and conserved R-parity due to the absence of
several final states in the GMSB case.

\section*{Acknowledgments}

This work was supported by ``Fonds zur F\"orderung der
wissenschaftlichen Forschung'' of Austria, project No.~P13139-PHY, by
Spanish DGICYT grants PB98-0693 and by the EEC under the TMR contract
HPRN-CT-2000-00148.  W.P. was supported by a fellowship from the
Spanish Ministry of Culture under the contract SB97-BU0475382.
D.R.~was supported by Colombian COLCIENCIAS fellowship.

\appendix

\section{Scalar Mass Matrices}

The mass matrix of the charged scalar sector follows from the
quadratic terms in the scalar potential~\cite{Akeroyd:1998iq,stop3}.
\begin{equation} 
V_{quadratic}=\mathbf{{S'}^-}
\mathbf{M_{S^{\pm}}^2}\mathbf{{S'}^+}
\label{eq:Vquadratic} 
\end{equation} 
where$\mathbf{{S'}^-}=[H_1^-,H_2^-,\tilde\tau_L^-,\tilde\tau_R^-]$.
For convenience reasons we will divide this $4\times4$ matrix into
$2\times2$ blocks in the following way:
\begin{equation} 
  \mathbf{M_{S^{\pm}}^2}=
  \left[
    \begin{array}{ccc}
      \mathbf{ M_{HH}^2} & \mathbf{ M_{H\tilde\tau}^{2}}^T \\ 
      \mathbf{ M_{H\tilde\tau}^2} & \mathbf{ M_{\tilde\tau\tilde\tau}^2} 
    \end{array}  
  \right]
  \label{eq:subdivM} 
\end{equation} 
where the charged Higgs block is 
\begin{eqnarray} 
  && \mathbf{ M_{HH}^2}= 
  \label{eq:subMHH} \\ \nonumber \\ 
  && \!\!\!\!\!\!
  \left[
    \begin{array}{ccc}
      B\mu{{v_2}\over{v_1}}+\quarter g^2(v_2^2-v_3^2)+\mu\epsilon_3 
      {{v_3}\over{v_1}}+\half h_{\tau}^2v_3^2+{{t_1}\over{v_1}} 
      & B\mu+\quarter g^2v_1v_2 \\
      B\mu+\quarter g^2v_1v_2 
      & B\mu{{v_1}\over{v_2}}+\quarter g^2(v_1^2+v_3^2)-B_3\epsilon_3 
      {{v_3}\over{v_2}}+{{t_2}\over{v_2}} 
    \end{array}
  \right]
  \nonumber 
\end{eqnarray} 
and $h_{\tau}$ is the tau Yukawa coupling. 
\begin{eqnarray} 
  && \mathbf{ M_{\tilde\tau\tilde\tau}^2}= 
  \label{eq:subtautau} \\ \nonumber \\ 
  && \!\!\!\!\!\!\!\!\!\!\!\!\!\!
  \left[
    \begin{array}{ccc}
      \half h_{\tau}^2v_1^2-\quarter g^2(v_1^2-v_2^2)+\mu\epsilon_3 
      {{v_1}\over{v_3}}-B_3\epsilon_3{{v_2}\over{v_3}}+{{t_3}\over{v_3}} 
      & {1\over{\sqrt{2}}}h_{\tau}(A_{\tau}v_1-\mu v_2) 
      \\ {1\over{\sqrt{2}}}h_{\tau}(A_{\tau}v_1-\mu v_2) 
      & m_{R_3}^2+\half h_{\tau}^2(v_1^2+v_3^2) 
      -\quarter g'^2(v_1^2-v_2^2+v_3^2) 
    \end{array}
  \right]
  \nonumber 
\end{eqnarray} 
The mixing between the charged Higgs sector and the stau sector is 
given by the following $2\times2$ block: 
\begin{equation} 
  \mathbf{ M_{H\tilde\tau}^2}=
  \left[
    \begin{array}{ccc}
      -\mu\epsilon_3-\half h_{\tau}^2v_1v_3+\quarter g^2v_1v_3 
      & -B_3\epsilon_3+\quarter g^2v_2v_3 
      \\ -{1\over{\sqrt{2}}}h_{\tau}(\epsilon_3v_2+A_{\tau}v_3) 
      & -{1\over{\sqrt{2}}}h_{\tau}(\mu v_3+\epsilon_3v_1) 
    \end{array}
  \right]
\label{eq:subHtau} 
\end{equation} 
As we see the charged Higgs bosons mix with charged sleptons.

In a similar way the real (imaginary) parts of the sneutrino mix the
scalar (pseudoscalar) Higgs bosons. The quadratic scalar potential
responsible for the neutral Higgs sector mass matrices includes
\begin{equation} 
  V_{quadratic}=\half\mathbf{{P'}^{0}}^T
  \mathbf{ M^2_{P^0}}\mathbf{{P'}^0}
  +\mathbf{{S'}^{0}}^T
  \mathbf{ M^2_{S^0}}\mathbf{{S'}^0}+\ldots
  \label{eq:NeutScalLag} 
\end{equation} 
where $\mathbf{{P'}^{0}}^T=[\varphi^0_1,\varphi^0_2,\tilde\nu_{\tau}^I]$, 
$\mathbf{{S'}^{0}}^T=\half[\chi^0_1,\chi^0_2,\tilde\nu_{\tau}^R]$ 
and the CP-odd neutral scalar mass matrix is
\begin{equation} 
  \mathbf{ M^2_{P^0}}=
  \left[
    \begin{array}{ccc}
      B\mu{{v_2}\over{v_1}}+\mu\epsilon_3{{v_3}\over{v_1}}+{{t_1}\over{v_1}} 
      & B\mu & -\mu\epsilon_3 \\ 
      B\mu & B\mu{{v_1}\over{v_2}}-B_3\epsilon_3{{v_3}\over{v_2}}
      +{{t_2}\over{v_2}} & -B_3\epsilon_3 \\ 
      -\mu\epsilon_3 & -B_3\epsilon_3 &  
      \mu\epsilon_3{{v_1}\over{v_3}}-B_3\epsilon_3{{v_2}\over{v_3}} 
      +{{t_3}\over{v_3}} 
    \end{array}
  \right]
\label{eq:neuscaM} 
\end{equation} 
The neutral CP-even scalar sector mass matrix in eq.~(\ref{eq:NeutScalLag}) 
is given by 
\begin{eqnarray}  
  &&\mathbf{ M_{S^0}^2}= 
  \label{eq:PseScalM} \\
  && \!\!\!\!\!\!\!\!\!\!\!\!\!\!\! 
  \left[
    \begin{array}{ccc}
      B\mu{{v_2}\over{v_1}}+\quarter g_Z^2v_1^2+\mu\epsilon_3 
      {{v_3}\over{v_1}}+{{t_1}\over{v_1}}  
      & -B\mu-\quarter g^2_Zv_1v_2  
      & -\mu\epsilon_3+\quarter g^2_Zv_1v_3  
      \\ -B\mu-\quarter g^2_Zv_1v_2  
      & B\mu{{v_1}\over{v_2}}+\quarter g^2_Zv_2^2-B_3\epsilon_3 
      {{v_3}\over{v_2}}+{{t_2}\over{v_2}}  
      & B_3\epsilon_3-\quarter g^2_Zv_2v_3  
      \\ -\mu\epsilon_3+\quarter g^2_Zv_1v_3  
      & B_3\epsilon_3-\quarter g^2_Zv_2v_3  
      & \mu\epsilon_3{{v_1}\over{v_3}}-B_3\epsilon_3{{v_2}\over{v_3}} 
      +\quarter g^2_Zv_3^2+{{t_3}\over{v_3}}  
    \end{array}
  \right] \nonumber 
\end{eqnarray} 
where we have defined $g_Z^2\equiv g^2+g'^2$. Note that, as a result
of CP invariance, the CP--even and CP--odd parts of the scalar mass
matrices are disjoint and do not mix with each other.
 
The three mass matrices in eqs.~(\ref{eq:subdivM}),
(\ref{eq:neuscaM}), and (\ref{eq:PseScalM}) are diagonalized by
rotation matrices which define the eigenvectors
\begin{equation} 
  \mathbf{S^+}=
  \mathbf{ R_{S^{\pm}}}\mathbf{{S'}^+}
  \,,\qquad\quad 
  \mathbf{P^0}= 
  \mathbf{ R_{P^0}}\mathbf{{P'}^0}
  \,,\qquad\quad 
  \mathbf{S^0}=
  \mathbf{ R_{S^0}}\mathbf{{S'}^0}\,, 
  \label{eq:eigenvectors} 
\end{equation} 
and the eigenvalues 
$\rm{diag}(0,m_{S^{\pm}_2}^2,m_{S^{\pm}_3}^2, 
m_{S^{\pm}_4}^2)=\mathbf{ R_{S^{\pm}}}\mathbf{ M_{S^{\pm}}^2} 
\mathbf{ R_{S^{\pm}}^T}$ for the charged scalars, \linebreak[4]
$\rm{diag}(0,m_{P^0_2}^2,m_{P^0_3}^2)$ $=\mathbf{ R_{P^0}} 
\mathbf{ M^2_{P^0}}\mathbf{ R_{P^0}^T}$ for the CP--odd neutral scalars,
 and 
$\rm{diag}(m_{S^0_1}^2,m{S^0_2}^2,m_{S^0_3}^2)$ $=\mathbf{ R_{S^0}} 
\mathbf{ M^2_{S^0}}\mathbf{ R_{S^0}^T}$ for the CP--even neutral scalars.
 
The matrices $R_{S^{\pm}}$, $R_{P^0}$ and $R_{S^0}$ specify the mixing
between the Higgs sector and the stau sector.

If a $3\times 3$ matrix $\mathbf{ M}$ has a zero eigenvalue, then the 
other two eigenvalues satisfy 
\begin{equation} 
m_{\pm}={1\over 2}{\rm Tr}\mathbf{ M} 
\pm {1\over2}\sqrt{\left({\rm Tr}\mathbf{ M}\right)^2 
-4(M_{11}M_{22}-M_{12}^2+M_{11}M_{33}-M_{13}^2+M_{22}M_{33}-M_{23}^2)} 
\label{eq:EigenExact}
\end{equation}
The CP-odd neutral scalar mass matrix eq.~(\ref{eq:neuscaM}) has a
zero determinant, so that its eigenvalues $m_{P^0_2}^2$ and
$m_{P^0_3}^2$ ($m_A^2$ and $m_{\tilde\nu_{\tau}^R}^2$ in the MSSM
limit) can be calculated exactly with the previous formula.

\end{document}